\documentclass[11pt,a4paper]{article}                           

\usepackage[bookmarksopen, bookmarksnumbered, bookmarksopenlevel=2, pdfencoding=auto]{hyperref}
\usepackage{tikz}
\usepackage{tikz-3dplot}
\usetikzlibrary{calc}
\usetikzlibrary{decorations} %
\usepackage[UKenglish]{babel}
\usepackage{amsmath, amscd}
\usepackage{amssymb}

\usepackage{hhline}
\usepackage[bf]{caption}
\usepackage{cite}
\usepackage[vcentermath]{youngtab}
\usepackage{array}

\usepackage{mathtools}

\usepackage{geometry}
\usepackage{multirow}

\makeatletter
\def\nobreakhline{%
  \noalign{\ifnum0=`}\fi
    \penalty\@M
    \futurelet\@let@token\LT@@nobreakhline}
\def\LT@@nobreakhline{%
  \ifx\@let@token\hline
    \global\let\@gtempa\@gobble
    \gdef\LT@sep{\penalty\@M\vskip\doublerulesep}
  \else
    \global\let\@gtempa\@empty
    \gdef\LT@sep{\penalty\@M\vskip-\arrayrulewidth}
  \fi
  \ifnum0=`{\fi}%
  \multispan\LT@cols
     \unskip\leaders\hrule\@height\arrayrulewidth\hfill\cr
  \noalign{\LT@sep}%
  \multispan\LT@cols
     \unskip\leaders\hrule\@height\arrayrulewidth\hfill\cr
  \noalign{\penalty\@M}%
  \@gtempa}
\makeatother
\hypersetup{
    pdftitle={G4-Flux and Standard Model Vacua in F-theory},
    pdfauthor={Ling Lin, Timo Weigand},
    pdfsubject={F-theory compactifications}
}
\numberwithin{equation}{section}
\numberwithin{table}{section}

\newcommand{\su}{\mathrm{u}}

\newcommand{\sv}{\mathrm{v}}
\newcommand{\sw}{\mathrm{w}}

\begin{document}

\baselineskip=14pt
\parskip 5pt plus 1pt 

\interfootnotelinepenalty=10000

\vspace*{-1.5cm}
\begin{flushright}    
  {\small

  }
\end{flushright}

\vspace*{2cm}
\begin{center}        
  {\LARGE $G_4$-Flux and Standard Model Vacua in F-theory}
\end{center}

\vspace*{0.75cm}
\begin{center}        
Ling Lin and Timo Weigand
\end{center}

\vspace*{0.15cm}
\begin{center}        
\emph{Institut f\"ur Theoretische Physik, Ruprecht-Karls-Universit\"at, \\
             Philosophenweg 19, 69120  Heidelberg, Germany}
\end{center}


\begin{abstract}

We study the geometry of gauge fluxes in four-dimensional F-theory vacua with gauge group $SU(3) \times SU(2) \times U(1) \times U(1)$ and its implications for phenomenology. 
The models are defined by a  previously introduced class of elliptic fibrations whose fibre is given as a cubic hypersurface in ${\rm Bl}_2{\mathbb P}^2$, with the non-abelian gauge group factors $SU(3) \times SU(2)$ engineered torically via the top construction.
To describe gauge fluxes on these fibrations we provide a classification of the primary vertical middle cohomology group in a fashion valid for 
any choice of base space. Using the ideal theoretic technique of primary decomposition we compute the cohomology classes of the matter surfaces associated with states charged under the non-abelian gauge group. These expressions allow us to interpret the cancellation of the pure and mixed non-abelian anomalies geometrically as a result of the general form of the matter surfaces, without reference to a specific type of gauge flux. 
Explicit results for the chiral indices of all matter states are obtained in terms of intersection numbers of the base and can be directly applied to any choice of base consistent with the fibration. As a demonstration we scan for globally consistent F-theory vacua on $\mathbb P^3$, ${\rm Bl}_1\mathbb P^3$ and ${\rm Bl}_2 \mathbb{P}^3$, and find a globally consistent flux configuration with the chiral Standard Model spectrum plus an extra triplet pair, which may be lifted by a recombination process.

\end{abstract}
 
\vspace*{\fill}

\thispagestyle{empty}
\clearpage
\setcounter{page}{1}


\newpage

\tableofcontents

\section{Introduction}

The structure of four-dimensional F-theory compactifications is enriched in two important aspects compared to their six-dimensional cousins:
The first is related to the appearance of cubic Yukawa couplings in the four-dimensional effective action. Geometrically such couplings 
are realised at fibral singularities over codimension-three points on the base ${\cal B}$ of the elliptically fibred Calabi--Yau fourfold $Y_4$.
The second novelty on Calabi--Yau fourfolds is the appearance of non-trivial gauge backgrounds. The latter are responsible for chirality in the massless spectrum of charged excitations and are therefore a crucial ingredient in the definition of a four-dimensional F-theory vacuum.
Under duality with M-theory the gauge background on a configuration of 7-branes on ${\cal B}$ maps to a background for the M-theory 3-form gauge potential $C_3$ and its field strength $G_4$ \cite{Becker:1996gj,Sethi:1996es,oai:arXiv.org:hep-th/9908088}. The full information about this background is captured by the so-called D\'eligne cohomology group $H^4_{\cal D}(Y_4, \mathbb Z(2))$ \cite{Donagi:1998vw,Curio:1998bva}. 
In order to determine the exact massless spectrum it is necessary to suitably parametrise elements of $H^4_{\cal D}(Y_4, \mathbb Z(2))$ in a way that preserves the information about the flat part of the gauge connections modulo gauge invariance \cite{Intriligator:2012ue,Bies:2014sra}.
By contrast, computing the chiral index of the charged spectrum merely requires keeping track of the gauge field strength, i.e.~the 4-form flux $G_4$. Such fluxes take values in the middle cohomology group $H^{2,2}(Y_4)$, which exhibits a remarkably rich structure by itself. Indeed, the middle cohomology decomposes into an orthogonal sum of three subspaces \cite{Greene:1993vm,Braun:2014xka},
\begin{align}
H^{(2,2)}(Y_4) = H^{(2,2)}_\text{\tiny hor}(Y_4) \oplus H^{(2,2)}_\text{\tiny vert}(Y_4) \oplus H^{(2,2)}_\text{\tiny rem}(Y_4).
\end{align}
The primary horizontal subspace $H^{(2,2)}_\text{\tiny hor}(Y_4)$ induces a superpotential on the space of complex structure deformations, but does not contribute to the chiral index of charged matter states. It has been studied from various perspectives in \cite{Greene:1993vm,Gukov:1999ya,Grimm:2009ef,Braun:2011zm,Krause:2012yh,Intriligator:2012ue,Bizet:2014uua,Lin:2015qsa,Jockers:2016bwi}.
The primary vertical subspace $H^{(2,2)}_\text{\tiny vert}(Y_4)$ induces a D-term potential for the K\"ahler moduli and indeed contributes to matter chirality. Finally, the remainder  $H^{(2,2)}_\text{\tiny rem}(Y_4)$ induces neither an F-term nor a D-term \cite{Braun:2014xka}. Its importance in F-theory model building is, for instance, owed to the fact \cite{Braun:2014xka} that it describes the so-called hypercharge flux \cite{Donagi:2008ca,Beasley:2008dc,Beasley:2008kw,Donagi:2008kj} in F-theory GUTs \cite{Mayrhofer:2013ara,Braun:2014pva}.

Our interest in this article is in the primary vertical subspace  $H^{(2,2)}_\text{\tiny vert}(Y_4)$. This space is generated by products of $(1,1)$-forms on $Y_4$. 
We will present an efficient method to explicitly parametrise this space on any elliptic fibration $Y_4$ whose fibre can be described as a hypersurface in a toric fibre ambient space. The space $H^{(2,2)}_\text{\tiny vert}(Y_4)$ follows from the space of products of $(1,1)$-forms by suitably implementing the ideal of linear relations. This method is particularly powerful if, as is the case for the fibrations of interest in this paper, the group of divisors on $Y_4$ coincides with the pullback of the divisor group on the full ambient space $X_{5}$. In this case all computations can be pushed onto the ambient space $X_5$. 
One of the most important points of this construction, which already underlied the analysis \cite{Lin:2015qsa}, is that it can be applied in a manner independent of the concrete choice of base space ${\cal B}$, provided ${\cal B}$ is compatible with the existence of a smooth fibration. The construction of vertical gauge fluxes can therefore be set up for a large family of F-theory compactifications with the same type of brane configuration as encoded in the fibration structure, but defined for different physical compactification manifolds, i.e.~base spaces. For each such specific base, additional linear relations between the divisor classes may arise, but this will merely lead to a further specialisation of the general expressions we find for generic base spaces. 
Examples of vertical F-theory fluxes without reference to a base have previously been studied in \cite{Braun:2011zm,Marsano:2011hv,Krause:2011xj} (including the first fully explicit three-generation GUT model in \cite{Krause:2011xj}), while a base independent classification in the spirit outlined above has been obtained for the $SU(n)$ Tate models with $n \leq 5$ in \cite{Krause:2012yh}.
References
\cite{Grimm:2011fx,Braun:2013yti,Cvetic:2013uta,Cvetic:2015txa} systematically construct the vertical gauge fluxes for various types of fibrations over specific choices of base spaces.


The chiral index of localised matter in representation ${\cal R}$ induced by a choice of vertical $G_4$-flux is computed by the intersection theoretic overlap \cite{Donagi:2008ca,Braun:2011zm,Marsano:2011hv,Krause:2011xj,Grimm:2011fx}
\begin{align}
\chi({\cal R}) = \int_{\gamma_{\cal R}} G_4
\end{align}
of $G_4$ with the matter surface $\gamma_{\cal R}$. We will use the methods of primary decomposition \cite{Cvetic:2013uta, Cvetic:2013jta, Lin:2014qga}
to explicitly compute the Poincar\'{e}-dual cohomology classes $[\gamma_{\cal R}]$
 as elements of $H^{(2,2)}_\text{\tiny vert}(Y_4)$. Again no reference to a concrete base space ${\cal B}$ will be made. This puts us in a position to explicitly determine the chiral index for any choice of vertical gauge flux over any base ${\cal B}$ as an integral over suitable forms on ${\cal B}$ alone. These general expressions can then easily be evaluated for a concrete choice of base space ${\cal B}$ compatible with the fibration. As an interesting consequence, we are able to argue for the cancellation of various types of gauge and discrete anomalies -- different from previous approaches -- without any explicit calculation of $G_4$. In fact, utilising our knowledge of the matter surfaces $\gamma_{\cal R}$, we can show for our class of fibrations that the transversality and quantisation condition on $G_4$ alone predict anomaly cancellation. Note that similar observations based on an analogous analysis were made in \cite{Lin:2015qsa}.
 
The general methods presented in this work are also applicable in phenomenological investigations. Concretely, we use them to study the class of fibrations introduced in \cite{Lin:2014qga} giving rise directly to the Standard Model gauge group\footnote{For simplicity we will not distinguish between the gauge algebra and the global gauge group \cite{Mayrhofer:2014opa} in this paper.} $SU(3) \times SU(2) \times U(1)_Y$ together with an extra $U(1)$ factor.
While the original motivation for F-theory model building was in the context of Grand Unified Models \cite{Donagi:2008ca,Beasley:2008dc,Beasley:2008kw,Donagi:2008kj,Weigand:2010wm,Maharana:2012tu}, F-theory also allows for a direct route to the Standard Model without an intermediate eight-dimensional GUT. Such models are particularly well motivated in the context of intermediate or high scale supersymmetry breaking, where the paradigm of gauge coupling unification may become less compelling. 
In this spirit \cite{Lin:2014qga} provided the first construction of F-theory fibrations with the Standard Model gauge group and matter representations by classifying 
all toric $SU(3) \times SU(2)$ tops on the elliptic fibrations of \cite{Borchmann:2013jwa,Cvetic:2013nia,Cvetic:2013uta,Borchmann:2013hta,Cvetic:2013jta} with gauge group $U(1)_1 \times U(1)_2$. 
Different realisations of the Standard Model gauge group in F-theory have been studied in \cite{Grassi:2014zxa,Klevers:2014bqa,Cvetic:2015txa,Halverson:2015jua}, including a three-generation model in \cite{Cvetic:2015txa}.

The approach of   \cite{Lin:2014qga} gives rise to five inequivalent types of fibrations, each with a variety of possible embeddings of the Standard Model hypercharge $U(1)_Y$ as a suitable linear combination of $U(1)_1$ and $U(1)_2$. Each of these embeddings in turn allows for  different identifications of the Standard Model matter content with the massless representations realised in the fibration. The pattern of Yukawa couplings as found geometrically is in agreement with the additional selection rule from the extra abelian gauge group factor and allows us in particular to distinguish between heavy and light families, depending on whether or not a Yukawa coupling with the Higgs field is present perturbatively (in the K\"ahler moduli).
While the extra selection rule does forbid some of the dimension four and five proton decay operators present a priori in the MSSM, typically the resulting models require intermediate scale supersymmetry breaking to guarantee sufficient stability of the proton, in agreement with the original scope of the construction.  

In this work we classify the vertical fluxes for one of the five Standard Model fibrations of  \cite{Lin:2014qga} and explicitly compute the chiral indices for generic base spaces. 
While the non-abelian anomalies are shown to be cancelled as a consequence of the form of the charged matter surfaces, verifying mixed abelian anomaly cancellation requires in addition the chiralities of certain charged singlet states which are harder to compute from first principles.
Reversing the logic, we instead employ the Green--Schwarz terms derived in \cite{Cvetic:2012xn} to determine expressions for these states that are valid for fibrations over any base $\cal B$.
These general results then form the starting point of a search for vacua whose chiral spectrum resembles as closely as possible that of the Standard Model. We restrict our search to the base spaces ${\cal B} \in \{ \mathbb P^3, {\rm Bl}_1\mathbb P^3,{\rm Bl}_2\mathbb P^3 \}$. For ${\cal B} = {\rm Bl}_1\mathbb P^3$ we find one fully consistent flux configuration which gives rise, at the level of chiral indices, to the exact Standard Model spectrum plus one extra pair of triplets. The latter can be removed from the massless spectrum by Higgsing the extra massless abelian gauge group. The concrete expressions for the chiralities  derived in a base independent manner allow for a systematic extension of our search to a multitude of more complicated base spaces. 

This article is organised as follows: In section \ref{sec_verticalfluxes} we summarise the description of vertical gauge fluxes on elliptic fibrations whose fibre is embedded into a  toric ambient space. The algorithm for determining a basis of $H^{(2,2)}_\text{\tiny vert}(Y_4)$ is described in appendix \ref{sec:calculating_flux_basis}.
In section \ref{sec_Anomalies} we apply these general considerations to the Standard Model fibrations of \cite{Lin:2014qga}. We determine the explicit form of the matter surfaces, with details relegated to appendix \ref{app:homology_classes}, and understand anomaly cancellation as a property of the Poincar\'{e}-dual cohomology classes in $H^{(2,2)}_\text{\tiny vert}(Y_4)$.
Our search algorithm for explicit three-generation models is the subject of section \ref{sec:pheno_part}, where we also present a benchmark model from our search on the example base ${\cal B} = {\rm Bl}_1\mathbb P^3$. Our conclusions are presented in section \ref{sec_concl}.


\section{Vertical \texorpdfstring{\boldmath $G_4$}{G4}-Flux in F-theory} \label{sec_verticalfluxes}

To set the stage we begin with a general description of $G_4$-fluxes in F-theory. We will be working on an elliptically fibred Calabi--Yau fourfold $Y_4$ given by
\begin{align}
\begin{CD}
		\mathbb{E}_\tau @>>> Y_4 \\
		@. 	@VV{\pi}V \\
		@. {\cal B} 
\end{CD}
\end{align}
whose base ${\cal B}$ is a K\"ahler threefold. Non-abelian gauge symmetry is associated with the fibre singularities in codimension-one, which we assume to be 
completely resolved. The exceptional divisors introduced in the process of this resolution will be collectively denoted by  ${\rm Ex}_i$. These are in one-to-one correspondence with the Cartan generators of the non-abelian gauge algebra.
We will furthermore assume the existence of a rational zero-section $S_0$. This assumption can, however, be dropped, and $G_4$-fluxes in fibrations without a zero-section have been analysed in \cite{Lin:2015qsa}. Since we are interested in models with two abelian gauge group factors, $Y_4$ is required to exhibit two extra rational sections $S_1$ and $S_2$. The generators of the associated abelian gauge groups via the Shioda homomorphism will be denoted by $\omega_i$.

\subsection[Generalities on \texorpdfstring{$G_4$}{G4}-Flux]{Generalities on \texorpdfstring{\boldmath $G_4$}{G4}-Flux}

The geometric data describing a consistent $G_4$-flux is an element of $H^{(2,2)}(Y_4)$ 
satisfying the transversality conditions
\begin{align}\label{eq:transversality_condition}
	\int_{Y_4} \, G_4 \wedge \pi^{-1} ( D^{( {\cal B} )}_a) \wedge \pi^{-1} (D^{( {\cal B} )}_b ) = \int_{Y_4} \, G_4 \wedge S_0 \wedge \pi^{-1} (D^{( {\cal B} )}_a)  = 0 \qquad \forall \, D^{( {\cal B} )}_a, D^{( {\cal B} )}_b \in H^{(1,1)}( {\cal B} ) \,.
\end{align}
These conditions are a formalisation of the requirement that the flux has `one leg along the fibre' \cite{oai:arXiv.org:hep-th/9908088} in order to dualise to a well-defined gauge flux in the F-theory limit.
In order for the non-abelian gauge symmetry in the F-theory limit to remain unbroken, we have to ensure in addition that
\begin{align}\label{eq:gauge_symmetry_condition}
	\int_{Y_4} \, G_4 \wedge {\rm Ex}_i \wedge \pi^{-1} (D^{( {\cal B} )}_{a}) = 0 \qquad \forall \, D^{( {\cal B} )}_a \in H^{(1,1)}( {\cal B} ).
\end{align}
The origin of this constraint will become clear in section (\ref{sec:determining_homology_class}).
If we are interested in the phenomenology of Standard-Model-like vacua, an extra restriction comes from requiring that the 
 hypercharge $U(1)_Y$  gauge potential does not receive a St\"uckelberg mass:
\begin{align}\label{eq:no_stueckelberg}
	\int_{Y_4} \, G_4 \wedge \omega_Y \wedge \pi^{-1} (D^{( {\cal B} )}_a) = 0 \qquad \forall \, D^{( {\cal B} )}_a \in H^{(1,1)} ( {\cal B} ) \, .
\end{align}
This condition depends on the particular choice of  linear combination 
\begin{align}
U(1)_Y = a \, U(1)_1 + b \, U(1)_2 \qquad  \longleftrightarrow \qquad \omega_Y  =  a \,  \omega_1 + b \, \omega_2,
\end{align}
where $\omega_1$ and $\omega_2$ are the generators of the two abelian gauge group factors.
In addition, any viable flux must obey the quantisation condition \cite{Witten:1996md}
\begin{align}\label{eq:quantisation_condition}
	G_4 + \frac{c_2(Y_4)}{2} \in H^4 (Y_4, \mathbb{Z}) \, .
\end{align}
Related to the quantisation condition is the cancellation of D3-tadpole: The number $n_3$ of D3-branes depends on the flux via \cite{Sethi:1996es}
\begin{align}\label{eq:tadpole_condition}
	n_3 = \frac{\chi(Y_4)}{24} - \frac{1}{2} \int_{Y_4} \, G_4 \wedge G_4 \, ,
\end{align}
with $\chi(Y_4)$ being the Euler number of the Calabi--Yau fourfold. Clearly $n_{3}$ needs to be an integer. It is typically assumed that an appropriately quantised flux will also lead to an integer D3-tadpole $n_3$.
To avoid anti-D3-branes, which would destabilise the compactification, we must require that 
 $n_3 \geq 0$.

Finally, as recalled already in the introduction, the middle cohomology of a fourfold splits into 
\begin{align}
H^{(2,2)}(Y_4) = H^{(2,2)}_\text{\tiny hor}(Y_4)  \oplus H^{(2,2)}_\text{\tiny vert}(Y_4)  \oplus H^{(2,2)}_\text{\tiny rem}(Y_4) .
\end{align}
 The primary horizontal component  $H^{(2,2)}_\text{\tiny hor}(Y_4)$ has been introduced  in \cite{Greene:1993vm} as the subspace which can be reached from $H^{(4,0)}(Y_4)$ by two successive variations of Hodge structure. Its elements are Poincar\'{e}-dual to those 4-cycles which are algebraic only on a subset of the complex structure moduli space.  The part which is computationally accessible in the most straightforward way is the primary vertical subspace, $H^{(2,2)}_\text{\tiny vert}(Y_4)$, which is generated by products of divisors. The dual 4-cycles are thus algebraic for every choice of complex structure moduli. The horizontal and the vertical subspaces are mapped onto each other by mirror symmetry \cite{Greene:1993vm,Grimm:2009ef,Bizet:2014uua} and are orthogonal with respect to the intersection pairing. The remainder $H^{(2,2)}_\text{\tiny rem}(Y_4)$ was introduced in \cite{Braun:2014xka} as the orthogonal complement of $H^{(2,2)}_\text{\tiny hor}(Y_4) \oplus H^{(2,2)}_\text{\tiny vert}(Y_4)$.
  In what follows, we will focus on the class of $G_4$-fluxes inside $H^{(2,2)}_\text{\tiny vert}(Y_4)$.  

\subsection{Vertical Flux on toric Hypersurfaces}\label{sec:vertical_fluxes}

Charged matter in some representation $\cal R$ localises on a curve $C_{\cal R} \subset  {\cal B} $ where two 7-branes intersect.
By M/F-theory duality, such matter originates  in M2-branes wrapping certain combinations of $\mathbb{P}^1$s fibred over $C_{\cal R}$. These fibrations form 4-cycles $\gamma_{\cal R} \subset Y_4$ called matter surfaces. The associated chiral index depends only on the cohomology class of the gauge background, i.e.~the flux $G_4$. It is 
given by integrating the flux over the corresponding matter surface $\gamma_{\cal R}$ \cite{Braun:2011zm,Marsano:2011hv,Krause:2011xj,Grimm:2011fx},
\begin{align}\label{eq:calculate_chiral_index}
	\chi({\cal R}) = \int_{\gamma_{\cal R}} \! G_4 = \int_{Y_4} \! G_4 \wedge [\gamma_{\cal R}] \, .
\end{align}
Here and in the following, we use the notation $[\gamma]$ to denote the homology class of the 4-cycle $\gamma$. By Poincar\'{e}-duality, $[\gamma]$ can be regarded as a 4-form in its own right. As we will explain in section \ref{sec:determining_homology_class}, most of our matter surfaces can be explicitly shown to have vertical homology classes $[\gamma]$. Therefore, their chiral indices are only affected by fluxes in the vertical cohomology $H_\text{vert}^{(2,2)}(Y_4)$.
By contrast, the exact vector-like spectrum, as opposed to the chiral index, is sensitive also to the gauge data not encoded in the flux $G_4$ alone. The missing information involves the intermediate Jacobian parametrising the flat 3-form connections on $Y_4$. A framework to extract this information for the computation of the exact massless matter spectrum is described in \cite{Bies:2014sra}. In the present paper we content ourselves with the computation of the chiral index and therefore focus on an efficient description of vertical $G_4$-fluxes.

By definition, the vertical cohomology is generated by products of divisors.\footnote{Throughout this paper, we will denote a divisor and its Poincar\'{e}-dual $(1,1)$-cohomology-form by the same variable.} 
On an elliptic fibration $Y_d \rightarrow  {\cal B} $ of complex dimension $d$ the Shioda--Tate--Wazir theorem \cite{Wazir:2001} states that all divisors are either pullbacks of divisors $D^{( {\cal B} )}_i$ in the base (`vertical divisors')\footnote{We will here and in the following use the same variable to denote divisors $D^{( {\cal B} )}$ of the base and the corresponding vertical divisors $\pi^{-1}(D^{( {\cal B} )})$.}, exceptional (blow-up) divisors, or sections of the fibration. 
In our constructions via tops, $Y_d$ is a hypersurface $\{P_T = 0\} \equiv \{P_T\}$ in an ambient space $X_{d+1} \rightarrow  {\cal B} $ of complex dimension $d+1$, which is the fibration of a toric variety -- the fibre ambient space -- over the same base $ {\cal B} $. 
Over each point in $ {\cal B} $, $\{P_T\}$ cuts out an elliptic curve inside the fibre ambient space. 
In particular, in the fibrations that will be considered in this paper, sections and exceptional divisors of $Y_d$ arise from restrictions $\{P_T \} \cap D^{(T)}_i$ of divisors $D^{(T)}_i \subset X_{d+1}$ defined by the top. Since $X_{d+1}$ and $Y_d$ also share the same base, they share the same vertical divisors, hence all divisors of $Y_d$ come from $X_{d+1}$ by restriction.

We can therefore reduce all the relevant computations involving vertical fluxes and matter surfaces to the intersection theory of divisors on the ambient space $X_{d+1}$. In appendix \ref{sec:calculating_flux_basis} we present a method to simplify such calculations for fibrations over a generic base $ {\cal B} $.
This is the technical foundation for the explicit classification of vertical gauge fluxes in the Standard Model fibrations of   \cite{Lin:2014qga}, to which we now turn.



\section{Classification of Fluxes in an F-theory (N)MSSM}\label{sec_Anomalies}



We now specialise the discussion to the elliptic fibrations  introduced in \cite{Lin:2014qga}. These define a class of toric F-theory compactifications with gauge group $SU(3) \times SU(2) \times U(1)^2$. 
The elliptically fibred Calabi--Yau fourfolds $Y_4 \rightarrow  {\cal B} $ are constructed as hypersurfaces in a ${\rm Bl}_2 \mathbb{P}^2$-fibration $X_5 \xrightarrow{\pi}  {\cal B} $ \cite{Borchmann:2013jwa,Cvetic:2013nia,Cvetic:2013uta,Borchmann:2013hta,Cvetic:2013jta}. The non-abelian gauge symmetry is realised torically using the technique of tops \cite{Candelas:1996su, Bouchard:2003bu}. 
For the case at hand there are 5 inequivalent tops -- labelled as ${\rm I} \times {\rm A}$, ${\rm I} \times {\rm B}$, ${\rm I} \times {\rm C}$, ${\rm III} \times {\rm A}$ and ${\rm III} \times {\rm B}$ in \cite{Lin:2014qga} -- giving rise to the Standard Model gauge algebra with a further $U(1)$. 
The $U(1)$s arise from a rank two Mordell--Weil group generated by sections with divisor classes $S_0$ (zero-section), $S_1$ and $U$. The non-abelian part of the gauge group is localised over two vertical divisors, $W_2 = [\{w_2\}]$ for $SU(2)$ and $W_3 = [\{w_3\}]$ for $SU(3)$. The tops define divisors $E_i \, (i=0,1)$ and $F_j \, (j = 0, 1, 2)$ with associated coordinates $e_i$ and $f_j$, respectively, such that 
\begin{align}
E_0 + E_1 = \pi^{-1}(W_2) \equiv W_2, \qquad  F_0 + F_1 + F_2 = \pi^{-1}(W_3) \equiv W_3.
\end{align}
 The intersections of the divisors $E_i$ and $F_j$ with the hypersurface $Y_4$ give rise to the exceptional divisors which resolve the non-abelian singularities; they are given by $\mathbb{P}^1$-fibrations over $W_2$ and~$W_3$, respectively. 
 The rational fibres are in one-to-one correspondence with the simple roots of $SU(2)$ and ~$SU(3)$ and can split into further $\mathbb{P}^1$s over matter curves and Yukawa points. 

{\renewcommand{\arraystretch}{1.1}
\begin{table}[ht]
	\begin{align*}
	\begin{array}{c | c || c c c c c | c c | c c c}
		\multicolumn{1}{c}{}& \multicolumn{1}{c||}{} & \multicolumn{10}{c}{\text{coordinates}} \\ \cline{3-12}
		\multicolumn{1}{c}{}& \multicolumn{1}{c||}{} & \su & \sv & \sw & s_0 & s_1 & e_0 & e_1 & f_0 & f_1 & f_2	\\ \hline \hline
		\multirow{4}{3.5em}{base divisor classes} & W_2 & \cdot & \cdot & \cdot & \cdot & \cdot & 1 & \cdot & \cdot & \cdot & \cdot \\
		& W_3 & \cdot & \cdot & \cdot & \cdot & \cdot & \cdot & \cdot & 1 & \cdot & \cdot \\
		& \alpha & \cdot & \cdot & 1 & \cdot & \cdot & \cdot & \cdot & \cdot & \cdot & \cdot \\
		& \beta & \cdot & 1 & \cdot & \cdot & \cdot & \cdot & \cdot & \cdot & \cdot & \cdot \\ \cline{1-12}
		\multirow{6}{3.5em}{fibre \& excep. divisors} & U & 1 & 1 & 1 & \cdot & \cdot & \cdot & \cdot & \cdot & \cdot & \cdot \\
		& S_0 & \cdot  & \cdot & 1 & 1 & \cdot & \cdot & \cdot & \cdot & \cdot & \cdot \\
		& S_1 & \cdot & 1 & \cdot & \cdot & 1 & \cdot & \cdot & \cdot & \cdot & \cdot \\ \cline{2-12}
		& E_1 & \cdot & \cdot & -1 & \cdot & \cdot & -1 & 1 & \cdot & \cdot & \cdot \\ \cline{2-12}
		& F_1 & \cdot & 1 & \cdot & \cdot & \cdot & \cdot & \cdot & -1 & 1 & \cdot \\
		& F_2 & \cdot & \cdot & -1 & \cdot & \cdot & \cdot & \cdot & -1  & \cdot & 1 \\ \hline \hline
		\multicolumn{1}{c}{}  & & -1 & 0 & 1 & -1 & 0 & 0 & 1 & 0 & 0 & 1 \\ 
		\multicolumn{2}{c||}{\text{top data}} & 1 & -1 & 0 & 0 & 1 & 0 & 0 & 0 & 1 & 0 \\
		\multicolumn{1}{c}{} & & \underline{0} & \underline{0} & \underline{0} & \underline{0} & \underline{0} & \underline{x} & \underline{x} & \underline{y} & \underline{y} & \underline{y}
	\end{array}
	\end{align*}
	\caption{Divisor classes and coordinates of the ambient space for model $\mathrm{I} \times \mathrm{A}$. The last row (`top data') describes (parts of) the fan of the ambient space $X_5$; for a specific base $ {\cal B} $ one has to fix the lattice coordinates $\underline{x}$ and $\underline{y}$ as well as further toric data completing the description of ${\cal B}$.}
	\label{tab:IxA-divisor-classes}
\end{table}}
Given a base $ {\cal B} $, the geometry is specified by a choice of divisor classes $W_2$ and $W_3$ as well as of two other base classes $\alpha, \beta$ which parametrise the ${\rm Bl}_2 \mathbb{P}^2$-fibration. In the following we will focus on one of the five models labelled ${\rm I} \times {\rm A}$. The fibration data is given by table \ref{tab:IxA-divisor-classes}.
The Stanley--Reisner ideal of the fibre ambient space depends on the triangulation of the top.\footnote{Of course all physical quantities in the F-theory limit are independent of the choice of triangulation.} As in \cite{Lin:2014qga} we choose a triangulation leading to the SR-ideal
\begin{align}\label{eq:SR-ideal-IxA}
\su \, \sv , \su \, \sw , \sw \, s_0 , \sv \, s_1 , s_0 \, s_1 , e_0 \, \sw , e_1 \, s_0 , e_1 \, \su , \, f_0\,\sw , f_0\,s_1 , f_1\,s_0 , f_1\,\sv , f_2\,s_0 , f_2\,s_1 , f_2\,\su , f_0\,e_1.
\end{align}
Furthermore one can read off the linear relations amongst the divisors from the non-trivial columns in table \ref{tab:IxA-divisor-classes}, e.g.~$[\sv] =  \beta + U + S_1 + F_1$. This leads to the following generators of the linear equivalence ideal,
\begin{align}\label{eq:LIN-generators}
{\rm LIN} = \langle \beta + U + S_1 + F_1 - [\sv] \, ,\, \, \alpha + U + S_0 - E_1 - F_2 - [\sw] \, , \, \, W_2 - E_0 - E_1 \, , \, \, W_3 - F_0 - F_1 - F_2 \, \rangle.
\end{align}
The polynomial
\begin{align}\label{eq:hypersurface_IxA}
  \begin{split}
    P_T =  &\sv \, \sw \, (c_{1;0,0}\,e_1 \, f_2 \, \sw \, s_1 + c_{2;,0,1} \, f_0 \, f_2 \, \sv \, s_0) + \su \, (b_{0;1,1}\,e_0 \, f_0 \, \sv^2 \, s_0^2 + b_1 \, \sv \, \sw \, s_0 \, s_1 + b_{2;0,0}\,e_1 \, f_1 \,f_2 \, \sw^2 \, s_1^2) + \\
    &\su^2 (d_{0;1,1}\,e_0 \, f_0 \, f_1\, \sv \, s_0^2 \, s_1 + d_{1;0,0} \,f_1 \, \sw \, s_0 \, s_1^2 + d_{2;1,1}\,e_0 \,f_0 \, f_1^2 \, \su \, s_0^2 \, s_1^2) 
  \end{split}
\end{align}
cuts out the Calabi--Yau hypersurface $Y_4$ with divisor class $[P_T] = [b_1] + U + [\sv] + [\sw] + S_0 + S_1$ in $X_5$. The coefficients are sections of specific line bundles, or -- equivalently -- transform as certain divisor classes,
\begin{align}\label{eq:sections_classes}
	\begin{split}
		& [b_{0;1,1}] = \alpha - \beta + \overline{\cal K} - W_2 - W_3 \, , \, \, [b_1] = \overline{\cal K} \, , \, \, [b_{2;0,0}] = \beta - \alpha + \overline{\cal K} \, ,\\
		& [c_{1;0,0}] = \overline{\cal K} - \alpha \, , \, \, [c_{2,0,1}] = \overline{\cal K} - \beta - W_3 \, , \\
		& [d_{0;1,1}] = \alpha + \overline{\cal K} - W_2 - W_3 \, , \, \, [d_{1;0,0}] = \beta + \overline{\cal K} \, , \, \, [d_{2;1,1}] = \alpha + \beta + \overline{\cal K} - W_2 - W_3 .
	\end{split}
\end{align}
Here $\overline{\cal K}$ is the anti-canonical class of the base $ {\cal B} $. In this model the $U(1)$-generators are
\begin{align}\label{eq:I-A-U(1)-generators}
 \begin{split}
    & \omega_1^{{\rm I} \times {\rm A}} \equiv \omega_1 = S_1 - S_0 - \overline{\cal K} +    \frac{1}{2} E_1 +  \frac{2}{3} F_1 + \frac{1}{3} F_2 ,\\
    & \omega_2^{{\rm I} \times {\rm A}} \equiv \omega_2 = U - S_0 - \overline{\cal K} - [c_{1;0,0}] + \frac{2}{3} F_1 + \frac{1}{3} F_2.
  \end{split}
\end{align}

This geometry gives rise to a rich spectrum of matter charged under the $SU(3) \times SU(2) \times U(1)^2$ gauge symmetry. The various matter representations ${\cal R}$ as well as the curves $C_{\cal R}$ on ${\cal B}$ over which this matter is localised are listed in table \ref{tab:spectrum_IxA}.

A cautionary remark is in oder: In the sequel we will derive explicit expressions e.g.~for the chiral indices of charged matter states in a manner which is formally independent of the specific choice of base space ${\cal B}$. This assumes, however, that the choice of base is compatible with the fibration structure. For instance, if the full ambient space $X_5$ allows for a toric description, then the cones of the toric fan describing $X_5$ must project to cones of the base ${\cal B}$. Furthermore, it must be checked that no singularities  of $X_5$ lie on the hypersurface $Y_4$.

\begin{table}[ht]
\begin{center}
  \begin{tabular}{c|c|c}
  	$\cal R$  & $U(1)$-charges & curve $C_{\cal R}$ in base \\ \hline\hline \rule{0pt}{3ex}
   $\mathbf{2}_1$ & $(\frac{1}{2}, -1)$ & $\{w_2\} \cap \{c_{2;0,1}\}$ \\ [.5ex] \hline \rule{0pt}{3ex}
   $\mathbf{2}_2$ & $(\frac{1}{2},1)$ & $\{w_2\} \cap \{ c_{1;0,0}^2 \, d_{1;0,0} - b_1 \, b_{2;0,0} \, c_{1;0,0} + b_{2;0,0}^2 \, c_{2;0,1} \, w_3 \}$ \\ [.5ex] \hline \rule{0pt}{3ex}
   \multirow{2}{*}{$\mathbf{2}_3$} & \multirow{2}{*}{$(\frac{1}{2}, 0)$} & $\{w_2\} \cap \{  b_{0;1,1}^2\,d_{1;0,0}^2 + b_{0;1,1}\,(b_1^2\,d_{2;1,1} - b_1\,d_{0;1,1}\,d_{1;0,0} -  2\,c_{2;0,1}\,d_{1;0,0}\,d_{2;1,1} \, w_3)$ \\
   & & $+ c_{2;0,1}\,w_3\, (d_{0;1,1}^2\,d_{1;0,0} - b_1\,d_{0;1,1}\,d_{2;1,1} + c_{2;0,1}\,d_{2;1,1}^2 \, w_3)\}$ \\ [.5ex] \hline \hline \rule{0pt}{3ex}
%
%
    $\mathbf{3}_1$ & $(\frac{2}{3}, -\frac{1}{3})$ & $\{w_3\} \cap \{b_{0;1,1}\}$ \\ [.5ex] \hline \rule{0pt}{3ex}
    $ \mathbf{3}_2$ & $(-\frac{1}{3}, -\frac{4}{3})$ & $\{w_3\} \cap \{c_{1;0,0}\}$ \\ [.5ex]\hline \rule{0pt}{3ex}
    $\mathbf{3}_3$ & $(-\frac{1}{3}, \frac{2}{3})$ & $\{w_3\} \cap \{b_{0;1,1}\,w_2\,c_{1;0,0} - b_1\,c_{2;0,1}\}$ \\ [.5ex]\hline \rule{0pt}{3ex}
    $\mathbf{3}_4$ & $(\frac{2}{3}, \frac{2}{3})$ & $\{w_3\} \cap \{b_1\,b_{2;0,0} - c_{1;0,0}\,d_{1;0,0}\}$ \\ [.5ex]\hline \rule{0pt}{3ex}
    $\mathbf{3}_5$ & $(-\frac{1}{3},-\frac{1}{3})$ & $\{w_3\} \cap \{b_{0;1,1}\,d_{1;0,0}^2 - b_1\,d_{0;1,1}\,d_{1;0,0} + b_1^2\,d_{2;1,1}\}$ \\ [.5ex] \hline \hline \rule{0pt}{3ex}
    $({\bf 3}, {\bf 2})$ & $(\frac{1}{6} , -\frac{1}{3})$ & $\{w_2\} \cap \{w_3\}$ \\ [.5ex] \hline\hline \rule{0pt}{3ex}
	${\bf 1}^{(1)}$ & $(1, -1)$ & $\{b_{0;1,1} \} \cap \{c_{2;0,1} \}$ \\ [.5ex] \hline \rule{0pt}{3ex}
	${\bf 1}^{(2)}$ & $(1, 0)$ & $C^{(2)}$ \\ [.5ex] \hline \rule{0pt}{3ex}
	${\bf 1}^{(3)}$ & $(1, 2)$ & $\{b_{2;0,0} \} \cap \{c_{1;0,0} \}$ \\ [.5ex] \hline \rule{0pt}{3ex}
	${\bf 1}^{(4)}$ & $(1, 1)$ & $C^{(4)}$ \\ [.5ex] \hline \rule{0pt}{3ex}
	${\bf 1}^{(5)}$ & $(0, 2)$ & $\{c_{1;0,0} \} \cap \{c_{2;0,1} \}$ \\ [.5ex] \hline \rule{0pt}{3ex}
	${\bf 1}^{(6)}$ & $(0, 1)$ & $C^{(6)}$
  \end{tabular}
\end{center}
\caption{Matter representations in the ${\rm I} \times {\rm A}$ model, together with the corresponding codimension 2 loci in $ {\cal B} $ over which they are localised. The singlet curves $C^{(2)}$, $C^{(4)}$ and $C^{(6)}$ cannot be written as complete intersections (see \cite{Lin:2014qga} for details).}
\label{tab:spectrum_IxA}
\end{table}

\subsubsection*{Fluxes over generic Bases}

To compute the vertical fluxes for a generic base $\cal B$ compatible with the fibration, we first need to construct the cohomology ring $H^{(k,k)}_\text{vert}(Y_4)$ in terms of a quotient ring. The technical procedure for this is summarised in appendix \ref{sec:calculating_flux_basis}. As a result we obtain a basis $\{t_i\}$ of $H^{(2,2)}_\text{vert}(Y_4)$. 
With the ansatz $G_4 = \lambda_i \, t_i$, the transversality and gauge symmetry conditions \eqref{eq:transversality_condition} and \eqref{eq:gauge_symmetry_condition} can be reduced -- as explained in the appendix -- to conditions of the form 
\begin{align}
	p_{abc}( \lambda_i) \int_{\cal B} D^{(\cal B)}_a \, D^{(\cal B)}_b \, D^{(\cal B)}_c = 0
\end{align}
in terms of intersection numbers on $\cal B$. The expressions $p_{abc}( \lambda_i)$ are linear in $\lambda_i$.
To ensure these conditions on any base $\cal B$, regardless of the precise form of the triple intersections $\int_{\cal B} D^{(\cal B)}_a \, D^{(\cal B)}_b \, D^{(\cal B)}_c$, the coefficients $p_{abc}( \lambda_i)$ must vanish individually. This leads to a set of independent equations linear in the $\lambda_i$. The non-trivial solutions give rise to the fluxes which are defined for any base. The same techniques have also been employed in \cite{Lin:2015qsa}. 
Previous classifications of vertical gauge fluxes over generic and concrete base spaces have been obtained in \cite{Krause:2012yh}  and  \cite{Grimm:2011fx,Braun:2013yti,Cvetic:2013uta,Cvetic:2015txa}, respectively.

For the ${\rm I} \times {\rm A}$ fibration, we obtain the following flux basis satisfying (\ref{eq:transversality_condition}) and (\ref{eq:gauge_symmetry_condition}):
\begin{align}\label{eq:general_flux_basis}
		& G^{z_1}_4 = -F_1\,(\beta + \overline{\cal K}) - (F_1 + 3\,S_1)\,W_3 + F_2\,(\beta + 3\,F_1 - 2\,\overline{\cal K} + W_3) \, , \notag \\
		& G^{z_2}_4 = (2\,F_1 - 2\,F_2)\,W_2 + E_1\,(6\,F_2 - 3\,W_3) \, , \notag\\
		& G^{z_3}_4 = 3\,F_2^2 + F_1\,(-\alpha - 2\,\overline{\cal K} + W_2) + F_2\,(-2\,\alpha + 3\,E_1 + 3\,F_1 - \overline{\cal K} - W_2 - 2\,W_3) + (2\,F_1 + 3\,S_1)\,W_3 \, , \notag \\
		& G^{z_4}_4 = E_1\,(3\,\beta + 6\,F_2 - 3\,\overline{\cal K} + 6\,S_1) + 2\,(F_1 - F_2 - 3\,S_1)\,W_2 \, , \notag \\
		& G^{z_5}_4 = S_1\,(E_1 + \overline{\cal K} + S_1 - W_2) \, ,\notag \\
		& G^{(i)}_4({\cal D}) = \omega_i \wedge {\cal D} \, \, \text{ for } i = 1, 2 \, \, \text{ and } \, \, {\cal D} \in H^{(1,1)}({\cal B}) \quad (U(1)_i\text{-fluxes}) \, .
\end{align}
The most general flux on a generic base ${\cal B}$ thus has the form 
\begin{align}
G_4 = \sum_i z_i\, G_4^{z_i} + G^{(1)}_4({\cal D}) + G^{(2)}_4({\cal D}').
\end{align}
The numerical coefficients $z_i \in \mathbb Q$ and the base divisor classes ${\cal D}$ and ${\cal D}'$ are subject to the quantisation condition (\ref{eq:quantisation_condition}).
Note that for an explicit choice of the fibration data and base ${\cal B}$, linear equivalences amongst the vertical divisors might render some of the above fluxes 
linearly dependent.
If no such linear dependences arise, one might wonder if additional fluxes can be constructed for a special base ${\cal B}$. However, it turns out that the only such fluxes are of the form $G^{(i)}_4({\cal D})$ for extra classes of ${\cal D}$ which may exist in addition to the generic base classes $\alpha, \beta, W_{2,3}, \overline{\cal K}$. In contrast, no additional fluxes of the form $G^{z_i}_4$ not related to a $U(1)_i$-flux can occur. This is of course under the assumption that the 
specific base ${\cal B}$ does not enforce further gauge enhancements, either non-abelian or abelian in the form of non-toric sections, on the full fibration. If this is case, the space of divisors on $Y_4$ and consequently also $H^{2,2}_{\rm vert}(Y_4)$ increases.

For completeness, we include here the D-terms  induced by the general flux for the  individual $U(1)$ gauge groups,
\begin{align}
	&\xi_1 \simeq  \int_{Y_4} G_4 \wedge \omega_1 \wedge J^{(\cal B)} = \int_{\cal B} J^{(\cal B)} \wedge \notag \\
	\begin{split}
		& \left( -2\,( {\cal D} + {\cal D}')\,\overline{\cal K} + \frac{1}{2} \, {\cal D}\,W_2 + \frac{1}{3}\, (2\,{\cal D} - {\cal D}') \,W_3 + 2\,\overline{\cal K}\,W_3\,z_1 - W_3^2\,z_1 - W_2\,W_3\,z_2 - 5\,\overline{\cal K}\,W_3\,z_3 \right.\\
	 	& + W_2\,W_3\,z_3 + 2\,W_3^2\,z_3 + \alpha\,({\cal D}' - W_3\,z_3) + 3\,\overline{\cal K}\,W_2\,z_4 - 4\,W_2\,W_3\,z_4 - \beta\,({\cal D}' + W_3\,z_1 + 3\,W_2\,z_4)  \\
 		& \left. + (\alpha - \beta + \overline{\cal K} - W_2 - W_3)\,(\beta - \overline{\cal K} + W_3)\,z_5 \vphantom{\frac{1}{2}} \right) \, ,
	\end{split}\notag \\
	&\xi_2 \simeq  \int_{Y_4} G_4 \wedge \omega_2 \wedge J^{(\cal B)} = \int_{\cal B} J^{(\cal B)} \wedge \label{eq:D-terms_generic_base} \\
	\begin{split}
		& \left( \alpha\,{\cal D} - \beta\,{\cal D} + 2\,\alpha\,{\cal D}' - {\cal D}\,\overline{\cal K} - 4\,{\cal D}'\,\overline{\cal K} - \frac{1}{3} \,{\cal D}\,W_3 + \frac{2}{3} \,{\cal D}'\,W_3 + 2\,\beta\,W_3\,z_1 - 4\,\overline{\cal K}\,W_3\,z_1 + 2\,W_3^2\,z_1 \right. \\
		& + 2\,W_2\,W_3\,z_2 - \alpha\,W_3\,z_3 - 3\,\beta\,W_3\,z_3 + \overline{\cal K}\,W_3\,z_3 +  W_2\,W_3\,z_3 - W_3^2\,z_3 + 6\,\beta\,W_2\,z_4 - 6\,\overline{\cal K}\,W_2\,z_4 \\
  		& \left. + 8\,W_2\,W_3\,z_4 + (\beta - \overline{\cal K} + W_3)\,(-\alpha + \beta - \overline{\cal K} + W_2 + W_3)\,z_5 \vphantom{\frac{1}{3}} \right) \, .
	\end{split} \notag	
\end{align}
Here $J^{(\cal B)}$ is the K\"ahler form on the base ${\cal B}$.
For an explicit realisation of the hypercharge generator $\omega_Y = \lambda_1 \, \omega_1 + \lambda_2 \, \omega_2$, the corresponding D-term must vanish in a phenomenologically viable model.
Likewise, consistency will require the D3-tadpole $n_3 = \chi(Y_4)/24 - 1/2\int_{Y_4} G_4^2$ to be integer; the expression for $1/2 \int_{Y_4} G_4^2$ is quite lengthy and its presentation is relegated to the appendix, cf.~formula \eqref{eq:general_D3-tadpole}.


\subsection{Homology Classes of Matter Surfaces}\label{sec:determining_homology_class}

A crucial input for computing the chiral spectrum are the homology classes of the matter surfaces $\gamma_{\cal R}$. To each representation $\cal R$, one associates $\dim {\cal R}$ different surfaces $\gamma^l$ which are fibrations of $\mathbb{P}^1$-chains $\Gamma^l_{\cal R}$ over the curve $C_{\cal R} \subset {\cal B}$.
\begin{table}[p!]
\begin{center}
  \begin{tabular}{c|c}
  	$\cal R$  & homology class $[\gamma_{\cal R}] = [P_T] \wedge [\tilde\gamma_{\cal R}]$ \\ \hline\hline \rule{0pt}{3ex}
   $\mathbf{2}_1$ & $c_{2;0,1} \, E_0 \, (b_1 + S_0 + U + \sv) \quad  =  \quad P_T \,\left\{ E_0 \, (\overline{\cal K} - S_1 - \beta - W_3) \right\}$ \\ [.5ex] \hline \rule{0pt}{3ex}
   \multirow{2}{*}{$\mathbf{2}_2$} & $E_0 \, \left[ (b_1 + b_{2;0,0} +S_0+\sv) \, (b_1 + S_1 + s_0 + U + \sv ) - 2\, \sv \, (d_{1;0,0} + S_0 + U) \right]$ \\ [1ex]
   & $= P_T \, \left\{ - \alpha \, E_0 - W_2 \, (F_1 - 2 \, \overline{\cal K} - S_0 + S_1 + U) + E_1 \, ( S_1 - F_2 - 2\, \overline{\cal K} + W_3)  \right\}$\\ [.5ex] \hline \rule{0pt}{3ex}
   \multirow{3}{*}{$\mathbf{2}_3$} & $ E_1 \, \left[ (2\,b_{0;1,1} + 2\,d_{1;0,0} ) \, (b_1  + \sw + \sv + S_1) + F_1 \, ( b_1 + S_1 ) + F_0 \, (b_1 + \sv ) \right. $ \\
   & $ - \left. (b_1  + \sv + S_1) \, (d_{0;1,1}+d_{1;0,0} +S_1 + 2\,F_1) + d_{1;0,0} \, (b_1 + S_1) \right] $ \\ [1ex]
   & $ = P_T \, \left\{ E_1 \, (\alpha - \beta + 2\, F_2 + 3\, \overline{\cal K} - 2\,S_1 - 2\,W_2 - 3\,W_3) + W_2 \, (\alpha - F_2 + S_0 + U) \right\} $ \\ [.5ex] \hline \hline \rule{0pt}{3ex}
    $\mathbf{3}_1$ & $F_1 \, \sw \, b_{0;1,1} \quad = \quad  P_T \, \left\{ F_2 \, (F_2 + E_1 - \alpha - W_3) + W_3 \, (\alpha - E_1 + S_0 + U) \right\}$ \\[.5ex] \hline \rule{0pt}{3ex}
    $\mathbf{3}_2$ & $F_0 \, U \, c_{1;0,0} \quad = \quad P_T \, \left\{ (F_1 + F_2 ) \, (\alpha - F_2) + W_3 \, (F_2 - \alpha - S_0) \right\} $ \\ [.5ex] \hline \rule{0pt}{3ex}
    $\mathbf{3}_3$ & $F_1 \, (b_1 + S_1 + \sw) \, (c_{2;0,1} + F_0) \quad  = \quad P_T \, \left\{ S_1 \, W_3 + F_1 \, (\overline{\cal K} - F_2 ) \right\} $ \\ [.5ex]\hline \rule{0pt}{3ex}
    \multirow{2}{*}{$\mathbf{3}_4$} & $F_0 \, (b_1 + U + S_0) \, (d_{1;0,0} + U + S_0)$ \\ [1ex]
    & $ = P_T \, \left\{ F_1 \, (\alpha - \beta - \overline{\cal K}) + F_2 \, (\alpha - \overline{\cal K} - F_2) + W_3 \, (F_2 - F_1 - S_1 - U - \alpha + \overline{\cal K}) \right\} $\\ [.5ex]\hline \rule{0pt}{3ex}
    \multirow{2}{*}{$\mathbf{3}_5$} & $F_2 \, \left[ (b_{0;1,1} + d_{1;0,0} +\sv) \, (b_{0;1,1} + 2\,\sv) - b_{0;1,1} \, (d_{0;1,1} + \sv ) - d_{2;1,1} \, \sv \right] $ \\ [1ex]
    & $ = P_T \, \left\{ F_2 \, (E_1 + 2\,F_1 + F_2 + \beta + \overline{\cal K} - W_2 - W_3 ) \right\}$ \\[.5ex] \hline \hline \rule{0pt}{3ex}
    $({\bf 3}, {\bf 2})$ & $E_0 \, F_2 \, (b_1 + \sv) \quad  = \quad P_T \, \left\{ E_0 \, F_2 \right\} $ \\ [.5ex] \hline\hline \rule{0pt}{3ex}
	\multirow{3}{*}{${\bf 1}^{(1)}$} & $b_{0;1,1} \, c_{2;0,1} \, (b_1+S_0 + U +\sv + \sw)$ \\ [1ex]
	& $ = P_T \, \left\{ (\overline{\cal K} - \beta) \, (\overline{\cal K} - \beta + \alpha) - S_1 \, (S_1 + E_1 + \overline{\cal K})  \right.$ \\
	& $ \hphantom{= P_T \, \{ \}} \qquad + \left. W_2 \, (\beta - \overline{\cal K} + S_1 + W_3) + W_3 \, (W_3 - \alpha + 2\, \beta - 2 \, \overline{\cal K} ) \right\} $\\ [.5ex] \hline \rule{0pt}{3ex}
	\multirow{3}{*}{${\bf 1}^{(3)}$} & $b_{2;0,0} \, c_{1;0,0} \, S_0$ \\ [1ex]
	& $ = P_T \, \left\{ S_0 \, \overline{\cal K} + S_1^2 - F_2 \, (F_1 + F_2) + \beta \, (S_1 - U) \right.  $ \\
	& $ \qquad \qquad \qquad \left.+ \alpha \, (F_1 + F_2 - S_0 + U) + W_3 \, ( F_2 - U - S_0 + S_1 - \alpha)  \right\}$ \\[.5ex] \hline \rule{0pt}{3ex}
	\multirow{2}{*}{${\bf 1}^{(5)}$} & $c_{1;0,0} \, c_{2;0,1} \, (b_1 + S_1 + S_0 + \sv + \sw)$ \\ [1ex]
	& $ = P_T \, \left\{ \overline{\cal K} \, (\overline{\cal K} - U) - S_1^2 - \beta \, (\overline{\cal K} + S_1 - U) - W_3 \, (\overline{\cal K} + S_1 - U) + \alpha \, (\beta - \overline{\cal K} + W_3) \right\} $ \\[.5ex] \hline\hline \rule{0pt}{3ex}
	\multirow{2}{*}{$\widetilde{{\bf 1}^{(2)}}$} & $ P_T \, \left\{ S_1 \, (3\,S_1 -2\,\alpha + 3\,\beta + 2\,E_1 - 2\,\overline{\cal K} + 4\,W_3) + U \, (\alpha - \beta + 2\,\overline{\cal K} - W_2 - 3\,W_3) \right. $\\
	& $\left. - F_2 \, (2\,E_1 + F_1 + 2\,F_2) \right\}$ \\[.5ex] \hline \rule{0pt}{3ex}
	\multirow{2}{*}{$\widetilde{ {\bf 1}^{(4)}}$} & $P_T \, \left\{ S_1 \, (2\,\alpha - 3\,S_1 - 4\,\beta - E_1 - 3\,\overline{\cal K} + W_2 - 2\,W_3) \right. $\\
	& $ \left. + U \, (2\,\beta - 2\,\alpha - 2\,\overline{\cal K} + W_2 + 4\,W_3) + F_2 \, (E_1 + 2\,F_1 + 3\,F_2) \right\}$ \\[.5ex] \hline \rule{0pt}{4ex}
	$\widetilde{ {\bf 1}^{(6)} }$ & $P_T \, \left\{ 2\,S_1\,(S_1 - \alpha + 2\,\beta + \overline{\cal K} + W_3) + U\,(-2\,\beta - 2\,\overline{\cal K} + W_2 - W_3) + (E_1 - F_2)\,F_2 \right\} $
  \end{tabular}
\end{center}
\caption{Homology classes of matter surfaces, given as 4-cycles in the ambient space and on the hypersurface. For formatting reasons we omit the square brackets indicating divisor classes of sections as well as wedge product symbols. The classes for the last three entries are not the actual matter surfaces, but give rise to the correct chiral index when integrated with a valid $G_4$ flux. See section \ref{sec:determining_singlet_classes} for more details.}
\label{tab:matter_surface_classes}
\end{table}
For matter in a non-trivial representation $\cal R$ under the non-abelian gauge group, which is localised on a divisor $W = [\{w\}]$, different weight states differ by linear combinations of simple roots. In homology, the difference $[\gamma^l] - [\gamma^k]$ for two different weights is therefore a linear combination $\sum_n \delta_n \, {\rm Ex}_n \wedge [p]$, where the numerical coefficients $\delta_n$ are dictated by representation theory, and $p$ defines the matter curve $C_{\cal R} = \{w\} \cap \{p\}$ (cf.~table \ref{tab:spectrum_IxA}). 
The condition (\ref{eq:gauge_symmetry_condition}) then ensures that for a valid flux $G_4$, the chirality within one representation is well-defined, 
\begin{align}
\int_{Y_4} G_4 \wedge [\gamma^l] - \int_{Y_4} G_4 \wedge [\gamma^k] = 0,
\end{align}
 i.e.~the flux does not break the non-abelian gauge symmetry in the F-theory limit. To keep things simple, we will therefore only refer to \textit{the} matter surface $\gamma$ of a representation $\cal R$, by which we mean the one (irreducible) surface given by the fibration of the $\mathbb{P}^1$ into which a root splits over $C_{\cal R}$; this $\mathbb{P}^1$ carries the weight charges of a state in $\cal R$ (which need not to be the highest weight). The splitting pattern of roots into various such weights can be encoded in so-called box graphs \cite{Hayashi:2014kca,Braun:2014kla}.

As explained in appendix \ref{app:homology_classes}, the matter surfaces have a natural description as algebraic 4-cycles in the ambient space $X_5$ in terms of some prime ideals.  Because not all $\gamma$'s are complete intersections, it requires some non-trivial polynomial algebra to determine the homology class of $\gamma$. At this point, we simply quote the results of this analysis in table \ref{tab:matter_surface_classes}, where we have listed all homology classes. The technical details underlying this method can be found in appendix \ref{app:homology_classes}. The resulting 4-cycle classes $[\gamma]$ allow for  the computation of the chiral index as 
\begin{align} \label{chiralR1}
\chi({\cal R}) = \int_{\gamma_{\cal R}} G_4 = \int_{X_5} \, G_4 \wedge [\gamma_{\cal R}],
\end{align}
 where we use the same notation $[\gamma_{\cal R}]$ for the Poincar\'{e}-dual 6-form in $X_5$ as for the Poincar\'{e}-dual 4-form on $Y_4$.

An important observation is that, on $X_5$, we can always find a `factorisation' $[\gamma] = [P_T] \wedge [\tilde{\gamma}]$, where the class $[\tilde\gamma]$ is a quadratic expression in the divisors. This means that on the hypersurface $\{P_T\}$, the homology of $\gamma$ is a vertical class (represented by the restriction of the cycle $\tilde\gamma$ to the hypersurface). The chiral index can then be re-expressed as $\chi = \int_{X_5} \, G_4 \wedge [P_T] \wedge [\tilde\gamma] \equiv \int_{Y_4} \, G_4 \wedge [\gamma]$. We will base our analysis of anomalies in the next section on the classes $[\tilde\gamma]$, which we have also included in table \ref{tab:matter_surface_classes}.

For completeness we note that in practise the method of appendix \ref{app:homology_classes} fails to determine the classes of the singlets ${\bf 1}^{(2)}$, ${\bf 1}^{(4)}$ and ${\bf 1}^{(6)}$. However, in the next section we will present an alternative approach using the anomaly conditions to find classes which at least yield the correct (i.e.~anomaly free) chirality. These classes are included in table \ref{tab:matter_surface_classes} for completeness.

With the knowledge of the matter surface classes, we can now compute the chiral indices (\ref{chiralR1}) induced by the fluxes \eqref{eq:general_flux_basis}. While in table \ref{tab:chiralities_missing_singlets}, we list the chiralities of the singlets ${\bf 1}^{(2)}$, ${\bf 1}^{(4)}$ and ${\bf 1}^{(6)}$, which are obtained from the results of section \ref{sec:determining_singlet_classes}, the chiralities of the other states can be readily computed and are shown in table \ref{tab:explicit_chiralities}.

Note that even though the specific basis (\ref{eq:general_flux_basis}) for the fluxes and the derivation of the matter surfaces made use of the choice of the SR-ideal (\ref{eq:SR-ideal-IxA}), all physical results are independent of the choice of triangulation of the fibre. 
In particular the results of tables \ref{tab:explicit_chiralities} and  \ref{tab:chiralities_missing_singlets}  for the chiral indices depend only on intersection numbers on the base. 
They can be applied straightforwardly to any choice of base ${\cal B}$ {\it provided} this choice gives rise to a consistent fibration structure in the sense specified in the paragraph after (\ref{eq:I-A-U(1)-generators}).

\begin{table}[p]
{
	\begin{align*}
		\begin{array}{@{}c |c}
			{\cal R} & \chi({\cal R}) \\ \hline \hline \rule{0pt}{3ex}
			\multirow{2}{*}{${\bf 2}_1$} & W_2\,( \overline{\cal K} - \beta - W_3)\, \left[ 3\,\beta\,z_4 + 3\,\overline{\cal K}\,z_4 - 6\,W_2\,z_4 + W_3\,(-3\,z_1 + 3\,z_2 + 3\,z_3 + 6\,z_4 - z_5) \right. \\
			& \left. + \alpha\,z_5 - \beta\,z_5 + \overline{\cal K}\,z_5 - W_2\,z_5 \right] -({\cal D} - 2\,{\cal D}' )\,W_2\,(\beta - \overline{\cal K} + W_3)/2 \\ [.5ex] \hline \rule{0pt}{3ex}
			{\bf 2}_2 & -3\,W_2\,  (\beta + \overline{\cal K})\, \left[(\beta - \overline{\cal K})\,z_4 + W_3\,(z_2 + 2\,z_4) \right] - ({\cal D} + 2\,{\cal D}')\,(2\,\alpha - \beta - 3\,\overline{\cal K})\,W_2/2\\ [.5ex] \hline \rule{0pt}{3.2ex}
			\multirow{3}{*}{${\bf 2}_3$} & W_2\, \left[ (\alpha - W_2 - W_3)\,W_3\,(3\,z_3 - z_5) + \overline{\cal K}^2\,(-6\,z_4 + z_5) + \beta^2\,(6\,z_4 + z_5) \right. \\
			& + \overline{\cal K}\, \left( W_3\,(6\,z_2 + 9\,z_3 + 12\,z_4 - 2\,z_5) + (\alpha - W_2)\,z_5) + \beta\,((-\alpha - 2\,\overline{\cal K} + W_2)\,z_5 \right. \\
  			& \left. \left. + W_3\,(6\,z_2 + 3\,z_3 + 12\,z_4 + 2\,z_5) \right) \vphantom{\overline{K}^2} \right] -{\cal D}\,W_2\,(-\alpha - 2\,\overline{\cal K} + W_2 + W_3) \\ [1ex] \hline\hline \rule{0pt}{3ex}
			\multirow{2}{*}{${\bf 3}_1$} & W_3\,(W_2 + W_3 -\alpha + \beta - \overline{\cal K} )\, \left[W_3\,z_1 - 2\,W_2\,z_2 + \alpha\,z_3 - W_2\,z_3 - 2\,W_3\,z_3 - 2\,W_2\,z_4 \right. \\
			& + \left. \overline{\cal K}\,(z_1 - z_3 - z_5) + W_3\,z_5 + \beta\,(z_1 + z_5) \right] + (2\,{\cal D} - {\cal D}')\,(\alpha - \beta + \overline{\cal K} - W_2 - W_3)\,W_3/3\\ [.5ex] \hline\rule{0pt}{3ex}
			\multirow{2}{*}{${\bf 3}_2$} & W_3 \, (\alpha - \overline{\cal K})\, \left[ \beta\,z_1 + W_3\,z_1 - 2\,W_2\,z_2 + \alpha\,z_3 - W_2\,z_3 - 2\,W_3\,z_3 + \overline{\cal K}\,(z_1 + 2\,z_3) - 2\,W_2\,z_4 \right] \\
			& + ({\cal D} + 4\,{\cal D}')\,(\alpha - \overline{\cal K})\,W_3 /3  \\ [.5ex] \hline\rule{0pt}{3ex}
			\multirow{4}{*}{${\bf 3}_3$} & W_3\, \left[\beta\,W_3\,(-z_1 + z_3 - 2\,z_5) - \beta\,\overline{\cal K}\,(z_1 + z_3 - 2\,z_5) + \beta^2\,(z_1 - z_5) + \alpha\,\overline{\cal K}\,(z_3 - z_5)  \right. \\ 
			& + W_3^2\,(-2\,z_1 + z_3 - z_5)  + \alpha\,\beta\,(z_3 + z_5) + \alpha\,W_3\,(z_3 + z_5) - 
  \overline{\cal K}^2\,(2\,z_1 + z_3 + z_5)  \\
  			& + \overline{\cal K}\,W_2\,(-2\,z_2 - z_3 + 4\,z_4 + z_5) - \beta\,W_2\,(2\,z_2 + z_3 + 8\,z_4 + z_5) - W_2\,W_3\,(2\,z_2 + z_3 + 8\,z_4 + z_5)\\
  			&  \left. + \overline{\cal K}\,W_3\,(5\,z_1 - 3\,z_3 + 2\,z_5) \right] + ({\cal D} - 2\,{\cal D}')\,W_3\,(\beta - 2\,\overline{\cal K} + W_3) /3\\ [.5ex] \hline \rule{0pt}{3ex}
			\multirow{3}{*}{${\bf 3}_4$} & W_3\, \left[ -\beta^2\,z_1 + \alpha\,\overline{\cal K}\,z_1 + \alpha\,W_3\,(z_1 - 2\,z_3) - \beta\,W_3\,(z_1 - 2\,z_3) - 2\,\overline{\cal K}\,W_3\,(z_1 - 2\,z_3) \right. \\
			& + \alpha\,\beta\,(z_1 - z_3) + \overline{\cal K}^2\,(z_1 - z_3) + \alpha^2\,z_3 + \beta\,\overline{\cal K}\,z_3 - 
  \alpha\,W_2\,(2\,z_2 + z_3 + 2\,z_4) \\
  			&\left. + \beta\,W_2\,(2\,z_2 + z_3 + 2\,z_4) + 2\,\overline{\cal K}\,W_2\,(2\,z_2 + z_3 + 2\,z_4) \right] -2\,({\cal D} + {\cal D}')\,(\alpha - \beta - 2\,\overline{\cal K})\,W_3/3 \\ [.5ex] \hline \rule{0pt}{3.2ex}
			\multirow{4}{*}{${\bf 3}_5$} & W_3\, \left[ - \beta^2\,z_1 - \alpha\,W_3\,z_1 - \alpha^2\,z_3 - \beta\,W_3\,z_3 - \alpha\,\beta\,(z_1 + z_3) + 3\,\overline{\cal K}^2\,(z_1 + z_3) + W_3^2\,(z_1 + z_3) \right. \\
			&  + \beta\,\overline{\cal K}\,(2\,z_1 + z_3) - \alpha\,\overline{\cal K}\,(z_1 + 2\,z_3) - 2\,\overline{\cal K}\,W_3\,(z_1 + 2\,z_3) + 2\,\alpha\,W_2\,(z_2 + z_3 + z_4) \\
			& - W_2^2\,(2\,z_2 + z_3 + 2\,z_4) + \beta\,W_2\,(z_1 + 2\,z_2 + z_3 + 2\,z_4) + \overline{\cal K}\,W_2\,(z_1 + 6\,z_2 + 2\,z_3 + 6\,z_4)\\
			&  \left. + W_2\,W_3\,(z_1 - 2\,(z_2 + z_4)) \vphantom{\overline{\cal K}^2} \right] -({\cal D} + {\cal D}')\,(\alpha + \beta + 3\,\overline{\cal K} - W_2 - W_3)\,W_3/3  \\ [1ex] \hline \hline \rule{0pt}{3ex}
			\multirow{2}{*}{$({\bf 3},{\bf 2})$} & W_2\,W_3\, \left[ -\alpha\,z_3 - \beta\,(z_1 - 3\,z_4) + W_2\,(2\,z_2 + z_3 + 2\,z_4) \right.  + W_3\,(-z_1 + 3\,z_2 + 2\,z_3 + 6\,z_4) \\
			&\left. - \overline{\cal K}\,(z_1 + 6\,z_2 + 2\,z_3 + 9\,z_4) \right]  + ({\cal D} - 2\,{\cal D}')\,W_2\,W_3/6\\ [.5ex] \hline \hline \rule{0pt}{3ex}
			\multirow{2}{*}{${\bf 1}^{(1)}$} & (-{\cal D} + {\cal D}')\,(\alpha - \beta + \overline{\cal K} - W_2 - W_3)\,(\beta - \overline{\cal K} + W_3) + (\overline{\cal K} - \beta - W_3)\,(-\alpha + \beta - \overline{\cal K} + W_2 + W_3) \\
			& \wedge \left[ W_2\,(6\,z_4 + z_5) - (\alpha - 2\,\beta + \overline{\cal K})\,z_5  + W_3\,(3\,z_1 - 3\,z_3 + 2\,z_5) \right] \\ [.5ex] \hline \rule{0pt}{3ex}
			{\bf 1}^{(3)} & ({\cal D} + 2\,{\cal D}')\,(\alpha - \overline{\cal K})\,(\alpha - \beta - \overline{\cal K}) \\ [.5ex] \hline \rule{0pt}{3ex}
			\multirow{2}{*}{${\bf 1}^{(5)}$} & (\alpha - \overline{\cal K})\,(\beta - \overline{\cal K} + W_3)\,(W_3\,(3\,z_1 - 3\,z_3 + z_5) + W_2\,(6\,z_4 + z_5) -(\alpha - \beta + \overline{\cal K})\,z_5) \\
			& + 2\,{\cal D}'\,(\alpha - \overline{\cal K})\,(\beta - \overline{\cal K} + W_3)
		\end{array}
	\end{align*}}
	\caption{Chiral indices of states with known matter surfaces in terms of the general flux basis \eqref{eq:general_flux_basis}.}\label{tab:explicit_chiralities}
\end{table}

\subsection{Gauge Anomaly Cancellation on generic Bases}\label{sec:gauge_anomalies}

In the presence of vertical $G_4$-flux the spectrum is chiral, leading to potential gauge anomalies in the 4D effective theory. By standard field theory reasoning the anomalies are given as sums of chiral indices, weighted with appropriate group theoretic factors. For our setup, where the gauge group is $SU(3) \times SU(2) \times U(1)_1 \times U(1)_2$, the possible types of non-trivial anomalies are $SU(3)^3$, $SU(n)^2-U(1)$, $U(1)_a - U(1)_b - U(1)_c$ and $U(1)-$gravitational. Here the $U(1)$s can be any linear combination $\lambda_1\,U(1) + \lambda_2\,U(1)_2$.

While pure non-abelian anomalies must vanish on their own\footnote{Geometrically, the vanishing of the pure non-abelian anomalies can be traced back to the possibility of redefining the affine node in an F-theory compactification \cite{Grimm:2015zea}.}, those involving $U(1)$s will in general require a Green--Schwarz (GS) mechanism to be cancelled. 
Whenever the GS-counterterms are non-zero, they will also lead to a flux-induced St\"uckelberg mass for the $U(1)$ gauge field. Applied to the hypercharge $U(1)_Y$ this means that all mixed anomalies involving $U(1)_Y$ must vanish by themselves.
Indeed, since we insist on a vanishing D-term (\ref{eq:no_stueckelberg}) to prevent a St\"uckelberg mass for $U(1)_Y$, the corresponding GS-counterterms are zero.

The form of the GS-counterterms in F-theory has been worked out in \cite{Cvetic:2012xn} via M/F-theory duality. Adapting these results to our notation and normalisation of $G_4$, we arrive at the following GS-counterterms for the corresponding anomalies:
\begin{align}
	SU(3)^3: & \quad 2\, \chi( {\bf 3}, {\bf 2} ) + \sum_i \chi({\bf 3}^{\rm A}_i) = 0 \label{eq:4D-anomalies-nonabelian} \\
	SU(3)^2 - U(1): & \quad  2 \, q ({\bf 3}, {\bf 2}) \, \chi({\bf 3}, {\bf 2}) + \sum_i q ({\bf 3}^{\rm A}_i) \chi({\bf 3}^{\rm A}_i) = - \int_{Y_4} \! G_4 \wedge \omega \wedge W_3 \label{eq:4D-anomalies-mixed-SU3}\\
	SU(2)^2 - U(1): & \quad  3 \, q ({\bf 3}, {\bf 2}) \, \chi({\bf 3}, {\bf 2}) + \sum_i q ({\bf 2}^{\rm I}_i) \chi({\bf 2}^{\rm I}_i) = - \int_{Y_4} \! G_4 \wedge \omega \wedge W_2 \label{eq:4D-anomalies-mixed-SU2}\\
	U(1)_a - U(1)_b - U(1)_c: & \quad   \sum_{\cal R} \dim({\cal R}) \, q_a ({\cal R}) \, q_b ({\cal R}) \, q_c ({\cal R}) \, \chi({\cal R}) = 3 \int_{Y_4} \! G_4 \wedge \pi_{*}( \omega_{( a} \wedge \omega_{\vphantom{(} b} ) \wedge \omega_{c)} \label{eq:4D-anomalies-u1cubed}\\
	U(1) - \text{gravitational}: & \quad  \sum_{\cal R} \dim({\cal R}) \, q ({\cal R}) \, \chi({\cal R}) = -6 \int_{Y_4} \! G_4 \wedge \overline{\cal K} \wedge \omega \, ,\label{eq:4D-anomalies-u1grav}
\end{align}
where on the left hand side, $q_{(\cdot)} ({\cal R})$ denotes the associated charge of the representation $\cal R$ under the $U(1)_{(\cdot)}$ generator $\omega_{(\cdot)} = \lambda^{(\cdot)}_1 \omega_1 + \lambda^{(\cdot)}_2 \omega_2$ (\ref{eq:I-A-U(1)-generators}).
Furthermore $\pi_{*}$ denotes the projection of 4-cycles in $Y_4$ to divisors of the base. The relevant values for us are
\begin{align}\label{eq:projection_of_omegas}
\begin{split}
	& \pi_* (\omega_1 \wedge \omega_1) = \frac{1}{2}\,W_2 + \frac{2}{3}\,W_3 - 2\,\overline{\cal K} \, , \\
	& \pi_* (\omega_1 \wedge \omega_2) = \pi_* (\omega_2 \wedge \omega_1) =  - \frac{1}{3} \, W_3 - \overline{\cal K} + \alpha - \beta \, , \\
	& \pi_* (\omega_2 \wedge \omega_2) = \frac{2}{3} \, W_3 - 4\,\overline{\cal K} + 2\,\alpha \, .
\end{split}
\end{align}
Finally note that in the chosen normalisation, the symmetrisation of the indices $(a,b,c)$ on the right hand side of (\ref{eq:4D-anomalies-u1cubed}) comes with a factor of $1/3!=1/6$.

It is straightforward, though tedious, to directly verify the matching \eqref{eq:4D-anomalies-nonabelian} -- \eqref{eq:4D-anomalies-mixed-SU2} of non-abelian anomalies with the chiralities in table \ref{tab:explicit_chiralities}.
Here, we present a different approach to anomaly cancellation in 4D F-theory, without any reference to the explicit form \eqref{eq:general_flux_basis} of the vertical fluxes. 
To this end, observe that both sides of the anomaly equations (\ref{eq:4D-anomalies-nonabelian}) -- (\ref{eq:4D-anomalies-u1grav}) can be regarded as integrals of $G_4$ over some 4-cycle. 
In particular, on the left hand sides the 4-cycle is a linear combination of matter surfaces. In the following we will show that these linear combinations  are such that they match their counter-part on the right hand side up to terms which are irrelevant for $G_4$-integration. In this sense, we will translate 4D anomaly cancellation into a geometric statement about the matter surfaces. Similar conclusions have been reached 
 in \cite{Lin:2015qsa} by studying 4D F-theory compactifications with $SU(5) \times U(1)$ and $SU(5) \times \mathbb{Z}_2$ symmetries.

\subsubsection{Non-abelian Anomalies}

We begin with the $SU(3)^3$ anomaly. According to (\ref{chiralR1}), the field theory expression (\ref{eq:4D-anomalies-nonabelian}) is calculated as $\int_{X_5} G_4 \wedge (2 \, [({\bf 3}, {\bf 2})] + \sum_i [{\bf 3}_i^{\rm A}] )$ in F-theory, where $[ {\cal R} ] \equiv [\gamma_{\cal R}]$ denotes the homology class of the matter surface associated to the state ${\cal R}$. With the homology classes explicitly given in table \ref{tab:matter_surface_classes}, it is now straightforward using the computational tools from appendix \ref{sec:calculating_flux_basis} to evaluate
\begin{align}\label{eq:su3-cubed-anomaly}
\begin{split}
	 2 \, [({\bf 3}, {\bf 2})] + \sum_i [{\bf 3}_i^{\rm A}] = &[P_T] \wedge \left\{ (2 \alpha - \beta - W_3) \wedge F_1 + (\alpha + \beta + W_2) \wedge F_2 - (E_1 + \alpha - \overline{\cal K}) \wedge W_3 \right\} \\
	  \equiv & [P_T] \wedge \eta_4 \\
	 \Longrightarrow &  \int_{X_5} G_4 \wedge \left( 2 \, [({\bf 3}, {\bf 2})] + \sum_i [{\bf 3}_i^{\rm A}] \right) = \int_{Y_4} G_4 \wedge \eta_4 \, .
\end{split}
\end{align}
The 4-form $\eta_4$ is obviously of the schematic form $ D_1^{({\cal B})} \wedge D_2^{({\cal B})} + \tilde{D}^{({\cal B})} \wedge {\rm Ex}_i$. Thus by construction, any $G_4$ satisfying (\ref{eq:transversality_condition}) and (\ref{eq:gauge_symmetry_condition}) leads to $\int_{Y_4} G_4 \wedge \eta_4 = 0$, i.e.~the $SU(3)^3$ anomaly is guaranteed to be cancelled for \textit{any} valid fluxes.

By analogous calculations, one finds for the $SU(3)^2 - U(1)$ anomaly with $U(1) = \lambda_1 \,U(1)_1 + \lambda_2 \,U(1)_2$ (cf.~table \ref{tab:spectrum_IxA} for the $U(1)$ charges) that
\begin{align}\label{eq:su3-u1-anomaly}
\begin{split}
	&2 \, q ({\bf 3}, {\bf 2}) \, [({\bf 3}, {\bf 2})] + \sum_i q ({\bf 3}^{\rm A}_i) \, [{\bf 3}^{\rm A}_i] = 2 \, \left( \lambda_1 \, q_1 + \lambda_2 \,q_2 \right)({\bf 3}, {\bf 2})  \, [({\bf 3}, {\bf 2})] + \sum_i \left( \lambda_1\, q_1 + \lambda_2\,q_2 \right) ({\bf 3}^{\rm A}_i) \, [{\bf 3}^{\rm A}_i] \\
	& = [P_T] \wedge \frac{1}{3} \left\{ \vphantom{\frac{1}{2}} \lambda_1 \left( (\alpha-2\beta - 3\overline{\cal K} - 2W_3) \wedge F_1 +(2W_2 -\alpha - \beta - 3\overline{\cal K}) \wedge F_2  + (\alpha - 2 E_1 +2 \overline{\cal K} ) \wedge W_3 \right) \right. \\
	&\hphantom{ = [P_T] \wedge \frac{1}{3} \left\{ \right. } + \lambda_2 \left( -2(\alpha + \beta + W_3) \wedge F_1 - (\alpha + \beta + 3\overline{\cal K} + W_2) \wedge F_2 + (\alpha + E_1 + 2\overline{\cal K} ) \wedge W_3 \right) \\
	& \hphantom{ = [P_T] \wedge \frac{1}{3} \left\{ \right. } \left. + (\lambda_1+\lambda_2) \, S_0 \wedge W_3 - (\lambda_1 \, S_1 +\lambda_2 \,U) \wedge W_3 \vphantom{\frac{1}{2}} \right\} \\
	&\equiv [P_T] \wedge \theta_4 \, .
\end{split}
\end{align}
Note that in $\theta_4$ the term $-(\lambda_1 \, S_1 + \lambda_2 \, U) \wedge W_3$ is the only one giving non-zero contributions when integrated with valid $G_4$ fluxes. Comparing with (\ref{eq:4D-anomalies-mixed-SU3}), we see that this precisely matches the GS-counterterm, which in this case is $- \int G_4 \wedge \omega \wedge W_3 = - \int G_4 \wedge (\lambda_1 \, S_1 + \lambda_2 \, U) \wedge W_3$.

Concerning the $SU(2)^2-U(1)$ anomaly, we have
\begin{align}\label{eq:su2-u1-anomaly}
\begin{split}
	& 3 q ({\bf 3}, {\bf 2}) [({\bf 3}, {\bf 2})] + \sum_i q ({\bf 2}^{\rm I}_i) \, [{\bf 2}^{\rm I}_i] \\
	& = [P_T] \wedge \left\{ \frac{\lambda_1}{2} \left( (2\alpha - W_2 - W_3) \wedge E_1 + (3\overline{\cal K} - \beta - W_2 - F_1 +S_0 - S_1 ) \wedge W_2 \right) \right. \\
	& \hphantom{= [P_T] \wedge \{} + \lambda_2 \left( (\alpha - \beta -\overline{\cal K} ) \wedge E_1 + (\beta - \alpha - F_1 - F_2 +\overline{\cal K} + W_3 + S_0 - U ) \wedge W_2 \right) \\
	& \hphantom{= [P_T] \wedge \{} \left. + (\lambda_1+\lambda_2) \,S_0 \wedge W_2 - (\lambda_1\,S_1 + \lambda_2\,U)\wedge W_2 \vphantom{\frac{a}{2}} \right\}
\end{split}
\end{align}
Again this result agrees with (\ref{eq:4D-anomalies-mixed-SU2}) by the same argument.

\subsubsection{Abelian Anomalies --- Determining Chiralities of missing Singlets}\label{sec:determining_singlet_classes}

The remaining types of chiral anomalies are $U(1)_a - U(1)_b - U(1)_c$ and $U(1)-$gravitational. Unfortunately, it is not possible to analyse these as above since the homology classes of the matter surfaces of the singlets ${\bf 1}^{(2)}, {\bf 1}^{(4)}$ and ${\bf 1}^{(6)}$, which contribute to said anomalies, are harder to determine. As a consequence, we are not able to demonstrate their cancellation directly.

But we can reverse the argumentation and use the anomaly matchings \eqref{eq:4D-anomalies-u1cubed} and \eqref{eq:4D-anomalies-u1grav} -- which we now assume to hold -- to determine the chiralities of those singlets. In fact, with the chiralities of all other states at hand (cf.~table \ref{tab:explicit_chiralities}), we can explicitly solve \eqref{eq:4D-anomalies-u1cubed} and \eqref{eq:4D-anomalies-u1grav} for $\chi ( {\bf 1}^{(i)})$, $i=2,4,6$, yielding the chiral indices as in table \ref{tab:chiralities_missing_singlets}. Similar analyses have been also performed e.g.~in \cite{Cvetic:2013uta}.

\begin{table}[ht]
\begin{align*}
	\begin{array}{c|c}
			{\cal R} & \chi( {\cal R} ) \\ \hline \hline \rule{0pt}{3.2ex}
			\multirow{7}{*}{${\bf 1}^{(2)}$} & 3\,\alpha^2\,W_3\,z_3 + 3\,W_3\, \left[ \vphantom{\overline{K}^2} (\beta^2 + \beta\,W_3 + \overline{\cal K}\,(-3\,\overline{\cal K} + 2\,W_3))\,z_1\right.  \\
			& \left. + (-2\,\beta^2 + 5\,\overline{\cal K}^2 - \beta\,(\overline{\cal K} + W_3) + (W_2 + W_3)^2 - \overline{\cal K}\,(3\,W_2 + 5\,W_3))\,z_3 \right] \\
			&+ 6\,(\beta + 2\,\overline{\cal K})\,W_2\,(\beta - \overline{\cal K} + W_3)\,z_4 - \alpha\,(2\,\beta + 4\,\overline{\cal K} - W_2 - W_3)\,(\beta - \overline{\cal K} + W_3)\,z_5 \\
			&+ (2\,\beta + 4\,\overline{\cal K} - W_2 - W_3)\,(\beta - \overline{\cal K} + W_3)\,(\beta - \overline{\cal K} + W_2 + W_3)\,z_5 \\
			& +  \alpha\,\left[ -3\,W_3\,(\beta - \overline{\cal K} + W_3)\,z_1+ 3\,(\beta + 2\,\overline{\cal K} - 2\,W_2 - W_3)\,W_3\,z_3 - 6\,W_2\,(\beta - \overline{\cal K} + W_3 )\,z_4 \right] \\
			& + {\cal D}\,\left[ \alpha^2 - 2\,\beta^2 + 5\,\overline{\cal K}^2 - 3\,\overline{\cal K}\,W_2 + W_2^2 + \alpha\,(\beta + 2\,\overline{\cal K} - 2\,W_2 - W_3) \right. \\
			& \left. - 5\,\overline{\cal K}\,W_3 + 2\,W_2\,W_3 + W_3^2 - \beta\,(\overline{\cal K} + W_3) \vphantom{\overline{K}^2} \right]\\ [1ex] \hline \rule{0pt}{3ex}
			\multirow{4}{*}{${\bf 1}^{(4)}$} & -3\,W_3\,\left[ \alpha^2 - \beta^2 + \alpha\,(\overline{\cal K} - W_2 - W_3) + \beta\,(-3\,\overline{\cal K} + W_2 + W_3) + 2\,\overline{\cal K}\,(-2\,\overline{\cal K} + W_2 + W_3) \right]\,z_3 \\
			& + (\overline{\cal K} -\alpha )\,(\beta - \overline{\cal K} + W_3)\,(-\alpha + \beta - \overline{\cal K} + W_2 + W_3)\,z_5 \\
			& + ({\cal D} + {\cal D}')\,\left[ -2\,\alpha^2 + \beta^2 + \beta\,(2\,\overline{\cal K} - W_2 - W_3) \right. \\
			& \left. + \overline{\cal K}\,(5\,\overline{\cal K} - 3\,(W_2 + W_3)) + \alpha\,(\beta - \overline{\cal K} + 2\,(W_2 + W_3)) \right] \\ [1ex] \hline \rule{0pt}{3.5ex}
			\multirow{6}{*}{${\bf 1}^{(6)}$} & 3\,W_3\, \left[ (\alpha^2 + \alpha\,\beta)\,z_3 +  (-\beta^2 + 5\,\overline{\cal K}^2 + \beta\,W_2 + W_3\,(W_2 + W_3) - \overline{\cal K}\,(W_2 + 4\,W_3))\,z_1 \right. \\ 
			& \left. \vphantom{\overline{K}^2} + \overline{\cal K}\,(-\beta - 3\,\overline{\cal K} + W_2 + W_3)\,z_3 \right] + 6\,W_2\,(\beta - \overline{\cal K} + W_3)\,(-\beta - 3\,\overline{\cal K} + W_2 + W_3)\,z_4 \\
			& - (2\,\beta + 4\,\overline{\cal K} - W_2 - W_3)\,(\beta - \overline{\cal K} + W_3)\,(\beta - \overline{\cal K} + W_2 + W_3)\,z_5 + \alpha \left\{ \vphantom{\beta^2} -3\,W_3\,(\beta - \overline{\cal K} + W_3)\,z_1 \right. \\
			& - 3\,W_3\,(-\beta - 2\,\overline{\cal K} + W_2 + W_3)\,z_3 - 6\,W_2\,(\beta - \overline{\cal K} + W_3)\,z_4 \\ \rule{0pt}{2.5ex}
			& \left. + (2\,\beta + 4\,\overline{\cal K} - W_2 - W_3)\,(\beta - \overline{\cal K} + W_3)\,z_5 \vphantom{\beta^2} \right\}	\\ \rule{0pt}{2.5ex}
			& + {\cal D}'\,(-2\,(\alpha^2 + \beta^2 - \alpha\,W_2 + \overline{\cal K}\,(-5\,\overline{\cal K} + 2\,W_2)) + (\alpha - \beta - 7\,\overline{\cal K} + W_2)\,W_3 + W_3^2)
	\end{array}
\end{align*}
\caption{Chiral indices of the singlets with unknown homology classes under the fluxes \eqref{eq:general_flux_basis}, computed by imposing anomaly cancellation.}\label{tab:chiralities_missing_singlets}
\end{table}

However, we can take further advantage of our knowledge of the matter surfaces of the other states in table \ref{tab:explicit_chiralities}. As we will show now, we can actually solve the $U(1)$-anomaly matchings \eqref{eq:4D-anomalies-u1cubed} and \eqref{eq:4D-anomalies-u1grav} at the level of matter surfaces. The result will be homology classes $[\widetilde{{\bf 1}^{(i)}}]$, $i=2,4,6$, which are valid for any base $\cal B$ and yield anomaly-free chiral indices $\chi( {\bf 1}^{(i)} ) = \int_{Y_4} G_4 \wedge [\widetilde{{\bf 1}^{(i)}}]$ for any $G_4$. Note that these classes will come in handy for our search of realistic chiral spectra in section \ref{sec:pheno_part}.

We first consider the $U(1)_1^2 - U(1)_2$ anomaly. By the charge assignments (\ref{tab:spectrum_IxA}) we see that, out of the missing singlets, only ${\bf 1}^{(4)}$ contributes to the left hand side of the anomaly matching (\ref{eq:4D-anomalies-u1cubed}). Assuming the matching to hold, we can therefore deduce that
\begin{align}\label{eq:matching-u1^2_u2}
\begin{split}
	& \int_{Y_4} \! G_4 \wedge [{\bf 1}^{(4)}] = \dim({\bf 1}^{(4)}) \, q_1^2 ({\bf 1}^{(4)}) \, q_2 ({\bf 1}^{(4)}) \, \chi ({\bf 1}^{(4)}) \\
	%
	%
	%
	%
	= & \int_{Y_4} \! G_4 \wedge \left( \frac{1}{2} \left( 4\,\pi_* (\omega_1 \wedge \omega_2) \wedge \omega_1 + 2\,\pi_* (\omega_1 \wedge \omega_1) \wedge \omega_2 \right)- \sum_{{\cal R} \neq {\bf 1}^{(4)}} \! \dim( {\cal R}) \, q_1^2 ( {\cal R}) \, q_2 ({\cal R}) \, [{\cal R}]  \right) \, .
\end{split}
\end{align}
Since this is now supposed to hold for any $G_4$-flux satisfying (\ref{eq:transversality_condition}) and (\ref{eq:gauge_symmetry_condition}), we conclude that we can compute the chirality of ${\bf 1}^{(4)}$ by integrating the flux over the 4-cycle
\begin{align}\label{eq:4cycle-u1^2_u2}
	\left( \frac{1}{2} \left( 4\,\pi_* (\omega_1 \wedge \omega_2) \wedge \omega_1 + 2\,\pi_* (\omega_1 \wedge \omega_1) \wedge \omega_2 \right)- \sum_{{\cal R} \neq {\bf 1}^{(4)}} \! \dim( {\cal R}) \, q_1^2 ( {\cal R}) \, q_2 ({\cal R}) \, [{\cal R}]  \right) \, .
\end{align}
Inserting all the relevant 4-cycle and divisor classes one finds a lengthy expression which we omit in the interest of readability. However, since terms of the form $D^{({\cal B})}_1 \wedge D^{({\cal B})}_2 + D^{({\cal B})} \wedge S_0 + D'^{({\cal B})} \wedge {\rm Ex}_i$ do not contribute to any $G_4$-integration, we can drop them for the purpose of computing chirality. The result is now much more compact,
\begin{align}\label{eq:singlet-class-1_4}
\begin{split}
	\widetilde{ [ {\bf 1}^{(4)} ] } & = S_1 \wedge (2\,\alpha - 3\,S_1 - 4\,\beta - E_1 - 3\,\overline{\cal K} + W_2 - 2\,W_3) + U \wedge (2\,\beta - 2\,\alpha - 2\,\overline{\cal K} + W_2 + 4\,W_3) \\
	& \hphantom{=} + F_2 \wedge (E_1 + 2\,F_1 + 3\,F_2) \, ,
\end{split}
\end{align}
while still giving the desired result $\chi( {\bf 1}^{(4)} ) = \int_{Y_4} G_4 \wedge \widetilde{ [{\bf 1}^{(4)}] }$. We stress that this is not the homology class of the actual matter surface of ${\bf 1}^{(4)}$.\footnote{\label{fncharges} For instance if one computed the Cartan charges of  ${\bf 1}^{(4)}$ based on this surface, one would find a non-zero result  $\int_{Y_4} {\rm Ex}_i \wedge \widetilde{ [ {\bf 1}^{(4)} ] } \wedge D^{({\cal B})} \neq 0$.}

As a first consistency check, we repeat the analogous computation for the $U(1)_1 - U(1)_2^2$ anomaly. Again, only ${\bf 1}^{(4)}$ out of the missing singlets contributes. Due to the charge assignments, we should now have
\begin{align*}
	\chi( {\bf 1}^{(4)} ) =  \int_{Y_4} \! G_4 \wedge \left( \frac{1}{2} \left( 4\,\pi_* (\omega_1 \wedge \omega_2) \wedge \omega_2 + 2\,\pi_* (\omega_2 \wedge \omega_2) \wedge \omega_1 \right)- \sum_{{\cal R} \neq {\bf 1}^{(4)}} \! \dim( {\cal R}) \, q_1 ( {\cal R}) \, q_2^2 ({\cal R}) \, [{\cal R}]  \right)  .
\end{align*}
Indeed, we find that while the 4-cycle inside the parentheses does not match the corresponding 4-cycle (\ref{eq:4cycle-u1^2_u2}) from the $U(1)_1^2 - U(1)_2$ anomaly, the difference is of the form $D^{({\cal B})}_1 \wedge D^{({\cal B})}_2 + D^{({\cal B})} \wedge S_0 + D'^{({\cal B})} \wedge {\rm Ex}_i$, i.e.~does not affect the calculation of the chiral index.

Having found a systematic way to compute the chirality for ${\bf 1}^{(4)}$, we can now use the $U(1)_1^3$ anomaly to pinpoint the homology class of ${\bf 1}^{(2)}$, since the other still unknown missing singlet ${\bf 1}^{(6)}$ does not contribute as it is not charged under $U(1)_1$. We proceed as before and isolate the chiral index to be determined from the matching condition (\ref{eq:4D-anomalies-u1cubed}):
\begin{align}\label{eq:matching-u1^3}
\begin{split}
	& \int_{Y_4} \! G_4 \wedge [{\bf 1}^{(2)}] = \dim ({\bf 1}^{(2)}) \, q_1^3 ({\bf 1}^{(2)}) \, \chi ({\bf 1}^{(2)}) =  \int_{Y_4} \! G_4 \wedge \left( \vphantom{\frac{1}{2}} \pi_* (\omega_1^2) \wedge \omega_1 - \sum_{ {\cal R} \neq {\bf 1}^{(2)} } \! \dim( {\cal R} ) \, q_1^3 ( {\cal R} ) \, [ {\cal R} ] \right) .
\end{split}
\end{align}
On the right hand side the sum now also runs over ${\cal R} = {\bf 1}^{(4)}$, for which we use the above result $\widetilde{ [ {\bf 1}^{(4)} ] }$ as the matter surface homology class. This yields
\begin{align}\label{eq:singlet-class-1_2}
	\begin{split}
		\widetilde{ [ {\bf 1}^{(2)} ] } & =  S_1 \wedge (3\,S_1 -2\,\alpha + 3\,\beta + 2\,E_1 - 2\,\overline{\cal K} + 4\,W_3) + U \wedge (\alpha - \beta + 2\,\overline{\cal K} - W_2 - 3\,W_3) \\
	& \hphantom{=} - F_2 \wedge (2\,E_1 + F_1 + 2\,F_2)
	\end{split}
\end{align}
up to terms of the form $D^{({\cal B})}_1 \wedge D^{({\cal B})}_2 + D^{({\cal B})} \wedge S_0 + D'^{({\cal B})} \wedge {\rm Ex}_i$.

Similarly, we can use the $U(1)_2^3$ anomaly to determine the corresponding 4-cycle for ${\bf 1}^{(6)}$. The matching condition in this case reads
\begin{align}\label{eq:matching-u2^3}
\begin{split}
	& \int_{Y_4} \! G_4 \wedge [{\bf 1}^{(6)}] = \dim ({\bf 1}^{(6)}) \, q_2^3 ({\bf 1}^{(6)}) \, \chi ({\bf 1}^{(6)}) =  \int_{Y_4} \! G_4 \wedge \left( \vphantom{\frac{1}{2}} \pi_* (\omega_2^2) \wedge \omega_2 - \sum_{ {\cal R} \neq {\bf 1}^{(6)} } \! \dim( {\cal R} ) \, q_2^3 ( {\cal R} ) \, [ {\cal R} ] \right) \, .
\end{split}
\end{align}
Using (\ref{eq:singlet-class-1_4}) for ${\cal R} = {\bf 1}^{(4)}$ (${\bf 1}^{(2)}$ does not contribute) we find
\begin{align}\label{eq:singlet-class-1_6}
	\widetilde{ [ {\bf 1}^{(6)} ] } = 2\,S_1\,(S_1 - \alpha + 2\,\beta + \overline{\cal K} + W_3) + U\,(-2\,\beta - 2\,\overline{\cal K} + W_2 - W_3) + (E_1 - F_2)\,F_2 \, .
\end{align}

As a further non-trivial consistency check, we consider the $U(1)-$gravitational anomalies (\ref{eq:4D-anomalies-u1grav}), now using the expressions $\widetilde{ [{\bf 1}^{(k)} ] }$ for $k = 2, 4, 6$ in the sum on the left hand side. With these we indeed verify that the 4-cycle $\sum_{\cal R} \dim( {\cal R} ) \, q_i( {\cal R}) \, [{\cal R}]$ is, up to terms of the form $D^{({\cal B})}_1 \wedge D^{({\cal B})}_2 + D^{({\cal B})} \wedge S_0 + D'^{({\cal B})} \wedge {\rm Ex}_k$, equal to $-6\,S_1 \wedge \overline{\cal K}$ for $i=1$ and $-6\,U \wedge \overline{\cal K}$ for $i=2$. This confirms that the matching of $U(1)-$gravitational anomalies is consistent with the chiralities for the missing singlets we deduced from the matching of $U(1)^3$ anomalies. Finally, we can compute the chiralities induced by the fluxes \eqref{eq:general_flux_basis}, which are identical to the results in table \ref{tab:chiralities_missing_singlets}.

Note again that the classes $\widetilde{ [{\bf 1}^{(k)} ] }$ for $k = 2, 4, 6$ are not the actual classes of the matter surfaces (see footnote \ref{fncharges}). 
This implies that we cannot make any statement about whether or not the matter surfaces of these states are vertical in homology, even though this is expected. 
It would require new techniques to address this issue.

\subsection{Cancellation of Witten Anomaly}\label{sec:witten_anomaly}

In addition to gauge anomalies analysed in the previous section, our model has a further source of perturbative anomaly: the famous Witten anomaly haunting $SU(2)$ gauge theories. Witten showed in \cite{Witten:1982fp} that a gauge theory with an $SU(2)$ gauge group must have an even number of doublets. In our model the statement can be phrased as
\begin{align}\label{eq:witten_anomaly}
	3 \, \chi( ({\bf 3}, {\bf 2}) ) + \sum_i \chi( {\bf 2}_i ) \equiv 0 \quad \text{mod} \, 2 \, .
\end{align}
In the following we will show that (\ref{eq:witten_anomaly}) is generically satisfied, \textit{assuming} that in a consistent fibration over a smooth base, an appropriately quantised flux always induces integer chiralities. Again we will only rely on the homology classes of matter surfaces and make not reference to any explicit $G_4$-fluxes.

First let us, similarly to the previous section, compute the 4-cycle contributing to the anomaly, 
\begin{align}\label{eq:witten_anomaly_cycle}
\begin{split}
	3\, [({\bf 3}, {\bf 2})] + \sum_i [ {\bf 2}_i ] = & [P_T] \wedge \left( -2\,E_1 \wedge F_2 - 2\, S_1 \wedge W_2 \vphantom{D^{({\cal B})}_a} \right. \\
	& \left. + \text{ terms of the form } D^{({\cal B})}_a \wedge D^{({\cal B})}_b + D^{({\cal B})} \wedge {\rm Ex}_i + D^{({\cal B})} \wedge S_0 \right ) \, .
\end{split}
\end{align}
For the Witten anomaly to be vanish, we thus need to show that 
\begin{align}\label{eq:witten_anomaly_cancellation_integerness}
\begin{split}
	\int_{Y_4} G_4 \wedge (-2\,E_1 \wedge & F_2 - 2\, S_1 \wedge W_2) = 2 \int_{Y_4} G_4 \wedge (-E_1 \wedge F_2 - S_1 \wedge W_2) \equiv 0 \quad \text{mod} \, 2 \\
	&\Longleftrightarrow \int_{Y_4} G_4 \wedge (-E_1 \wedge F_2 - S_1 \wedge W_2) \in \mathbb{Z} \, .
\end{split}
\end{align}
At this point we invoke the quantisation condition: Using the obvious fact that $E_1 \wedge F_2 + S_1 \wedge W_2$ is a manifestly integer class, (\ref{eq:quantisation_condition}) implies 
\begin{align}
	\int_{Y_4} \left(G_4 + \frac{1}{2} \, c_2(Y_4) \right) \wedge ( - E_1 \wedge F_2 - S_1 \wedge W_2 ) \in \mathbb{Z} \, ,
\end{align}
where the second Chern class $c_2 (Y_4)$ can be easily computed by adjunction (for the explicit expression see (\ref{eq:second-chern-class})). Thus (\ref{eq:witten_anomaly_cancellation_integerness}) follows if we can show that $\frac{1}{2} \int_{Y_4} c_2(Y_4) \wedge ( - E_1 \wedge F_2 - S_1 \wedge W_2 ) \in \mathbb{Z}$.
By straightforward calculation,
\begin{align}\label{eq:witten_anomaly_integral_c2}
\begin{split}
	&\frac{1}{2} \int_{Y_4} c_2(Y_4) \wedge ( - E_1 \wedge F_2 - S_1 \wedge W_2 ) =\\
	&\int_{\cal B} \left( \frac{1}{2} \, (W_2\,W_3^2 -  W_2^2\,W_3  +  W_2\,W_3\, \overline{\cal K} )+ \frac{1}{2}\,W_2\, ( \overline{\cal K}^2  - \,c_2({\cal B})) + \text{ integer terms} \right)
\end{split}
\end{align}
is not manifestly integer. However, it was shown in \cite{Collinucci:2010gz} that $c_2({\cal B}) - \overline{\cal K}^2$ is an even class for smooth complex threefolds. Thus the second summand is integer for a smooth base ${\cal B}$. To argue that the first term is also integer, we have to make use of our assumption that, for consistent geometries, all chiral indices are integer. If this is true, then, again by the quantisation condition, considering the class $[( {\bf 3}, {\bf 2}) ]$ (which is manifestly integer as a matter surface) yields the statement
\begin{align}\label{eq:witten_anomaly_integral_bifundamental}
\begin{split}
	\quad \chi( ( {\bf 3}, {\bf 2}) ) + \frac{1}{2} \int_{Y_4} c_2(Y_4) \wedge [( {\bf 3}, {\bf 2}) ] = \int_{Y_4} \left( G_4 + \frac{1}{2} \, c_2 (Y_4) \right) \wedge [( {\bf 3}, {\bf 2}) ] & \in \mathbb{Z}\\
	\overset{\chi(( {\bf 3}, {\bf 2})) \in \mathbb{Z}}{\Longrightarrow} \quad \int_{Y_4} \frac{1}{2} c_2(Y_4) \wedge [( {\bf 3}, {\bf 2}) ] = \int_{\cal B} \left( \frac{1}{2} (W_2^2\,W_3 + W_2\,W_3^2 - W_2\,W_3\,\overline{\cal K}) \right) & \in \mathbb{Z} \, ,
\end{split}
\end{align}
which then implies the integrality of (\ref{eq:witten_anomaly_integral_c2}).

To summarise: Based on the assumption that a consistent fibration implies integral chiral indices from a properly quantised flux, we have shown the cancellation of the Witten anomaly (\ref{eq:witten_anomaly}) for (consistent) fibrations \textit{over any (smooth) base} ${\cal B}$. Note that these assumptions were also crucial to show the cancellation of anomalies involving discrete symmetries, as presented in \cite{Lin:2015qsa}.


\section{Search for Realistic Models}\label{sec:pheno_part}

In this section we make contact with the phenomenological aspects of the fibrations of \cite{Lin:2014qga}. Our aim is to study whether, in explicit compactifications, $G_4$-fluxes can induce a realistic chiral spectrum in our F-theory `Standard Models'. 
To compare to realistic particle physics models, we first have to interpret the geometrically realised matter as Standard Model states. In \cite{Lin:2014qga} we have classified possible matchings of the geometric spectrum with the (N)MSSM based on the $U(1)$-charges of the states. For details we refer to section 6 therein. Note that for the model ${\rm I} \times {\rm A}$ there are three possibilities for the geometric $U(1)$'s to form the hypercharge $U(1)_Y = a\,U(1)_{1} + b\,U(1)_2$, namely $(a,b) = (1,0)$, $(a,b) = (0, -1/2)$ and $(a,b) = (-1, -1)$. These three possibilities lead to different $G_4$-solutions, as we require the flux to induce no D-term potential for hypercharge, see (\ref{eq:no_stueckelberg}).

To obtain explicit chiral indices we have to specify the full fibration data, i.e.~a choice for the base ${\cal B}$ and the classes $\alpha, \beta$ as well as $W_2, W_3$ entering (\ref{eq:sections_classes}). We then determine the space of valid $G_4$-fluxes and scan over part of it to search for configurations giving rise to realistic chiralities. In this scan we restrict ourselves to fluxes with induced chiral spectra in the range  $| \chi | <10$.

There are two possible routes one can take for such a search. With the results from the previous section, the obvious procedure would be to use the fluxes \eqref{eq:general_flux_basis} derived for a generic base, specialise to a concrete (consistent) fibration, and make use of the chiralities in tables \ref{tab:explicit_chiralities} and \ref{tab:chiralities_missing_singlets} as well as the formulae for the D-terms \eqref{eq:D-terms_generic_base} and the D3-tadpole \eqref{eq:general_D3-tadpole}. 
This route seems very attractive because one can impose the chirality of many states to take a desired value and then solve for the flux parameters $z_i$ and $\cal D, D'$. In particular, one could in principle pick one's favourite Standard Model identification from \cite{Lin:2014qga} and try to construct a suitable flux.
However, while the chiralities can often be tweaked into a more or less favourable scenario, we found that with this approach, it is generically very hard to find an appropriately quantised flux, e.g.~such that the D3-tadpole is integer. The existence of suitably quantised flux solutions which give rise to a given spectrum depends of course on the concrete choice of base ${\cal B}$, and for suitable ${\cal B}$ this approach may well lead to satisfactory results. 

In the sequel, we will follow an alternative strategy and instead scan over part of the flux landscape to investigate how closely the resulting models resemble the Standard Model. 
In principle one could use the basis \eqref{eq:general_flux_basis} and simply specialise it to a concrete base space ${\cal B}$. 
However, the lattice spanned by the fluxes \eqref{eq:general_flux_basis} is usually too coarse because the vertical divisors $\overline{\cal K}, \alpha$ and $\beta$ are in general not prime divisors. 
The effect is that the resulting chiral indices in tables \ref{tab:explicit_chiralities} and \ref{tab:chiralities_missing_singlets} are generically very large for order 1 values of the coefficients $z_i$, and  a suitable scan would require highly fractional coefficients, which in turn obscure the quantisation of the fluxes.
It is therefore more convenient to compute a basis of fluxes for each individual fibration. This basis will still be equivalent to the generic fluxes \eqref{eq:general_flux_basis} (modulo redundancies from the specialisation of the fibration) as a $\mathbb{Q}$-vector basis. 
However, we find that in general, these basis elements span a finer lattice in the sense that they induce small chiral indices even if we allow for integer coefficients.
This allows in particular for a finer scan over the flux landscape than using the fluxes \eqref{eq:general_flux_basis}. 

\subsection{Search Algorithm}

We have constructed fibrations over the toric bases ${\cal B} \in \{ \mathbb{P}^3, {\rm Bl}_1 \mathbb{P}^3 , {\rm Bl}_2 \mathbb{P}^3 \}$. For simplicity we identify each of the coordinates $w_2$ and $w_3$ describing the $SU(2)$ and $SU(3)$ brane divisors with one of the homogeneous coordinates of ${\cal B}$.\footnote{Any other more complicated identification requires working with complete-intersection fourfolds.} 
Having fixed this choice, we then restrict the classes $\alpha$ and $\beta$ such that all the sections (\ref{eq:sections_classes}) have effective classes.
 Each allowed pair $(\alpha, \beta)$ fixes a polytope for the toric ambient space $X_5$. To fully define $X_5$, we need to find a suitable triangulation of the polytope that defines a toric fan compatible with the fibration structure, and ultimately also determines the Stanley--Reisner-ideal. We use the \texttt{Sage} package \texttt{Topcom} to find all possible triangulations and then pick one whose SR-ideal contains (\ref{eq:SR-ideal-IxA}) as a subset. This allows us to use the results on the matter surfaces as listed in table \ref{tab:matter_surface_classes}, which crucially depend on the SR-ideal. 

Note that while for the bases $\mathbb{P}^3$ and ${\rm Bl}_1 \mathbb{P}^3$ it is always possible to find such a triangulation, this need not generally be the case. 
In such a situation, one would need to repeat the analysis of matter surfaces in section \ref{sec_Anomalies} with another suitable SR-ideal.
In our search we encounter this situation only for fibrations over the base ${\cal B} = {\rm Bl}_2 \mathbb{P}^3$. These particular models would not be suitable for phenomenological applications anyway, because they are only compatible with a fibration in which the divisors $W_2$ and $W_3$ of our fixed choice do not intersect on $\cal B$. On the resulting fourfold we would have no bifundamental $({\bf 3}, {\bf 2})$ states.

Having fully defined the toric ambient space $X_5$ it is straightforward to compute the (rational) cohomology ring (\ref{eq:cohom_ambient_space_general}) using \texttt{Sage}. Note that for toric spaces the vertical cohomology (\ref{eq:cohom_ambient_space_general}) constitutes in fact the full cohomology ring. This is of course not the case for the hypersurface $Y_4$. We then proceed to find a basis $\{t_k\}$ of $H^{(2,2)}_\text{vert}(Y_4, \mathbb{Q})$.
It is not necessarily the same as the basis of $H^{(2,2)}(X_5, \mathbb{Q})$, since different $(2,2)$-forms can  -- and in fact do -- become equivalent when restricted to the hypersurface $P_T$.\footnote{The inverse phenomenon would arise e.g.~when an ambient divisor splits into two independent divisors on the hypersurface. In the fibrations under consideration in this paper this does not occur. } 
As explained in the appendix, we use \texttt{Singular} to determine the basis $\{t_k\}$. The output is of the form $t_k = D_{a_k} \wedge D_{b_k}$, where $D_{a_k,b_k}$ are toric divisors of the ambient space $X_5$.

Postponing the question of quantisation, valid $G_4$-fluxes are linear combinations of $t_k$ that satisfy (\ref{eq:transversality_condition}), (\ref{eq:gauge_symmetry_condition}) and (\ref{eq:no_stueckelberg}).\footnote{Note again there are three different choices for the hypercharge that lead to three inequivalent sets of valid $G_4$-fluxes.} 
We add one further restriction on the fluxes, namely that the chirality of the bifundamental states $({\bf 3}, {\bf 2})$ is $\chi(({\bf 3}, {\bf 2})) = 3$. This has obvious phenomenological motivation as we only have one matter curve hosting this representation, and thus all three generations of left-handed quarks must reside here. We accommodate this constraint in our search by determining the subspace $V \subset H^{(2,2)}_\text{vert}(Y_4, \mathbb{Q})$ satisfying
\begin{align}\label{eq:subspace_V}
\begin{split}
	\int_{Y_4} v \wedge D^{({\cal B})}_a \wedge D^{({\cal B})}_b = \int_{Y_4} v \wedge Z \wedge D^{({\cal B})}_a = \int_{Y_4} v \wedge \omega_Y \wedge D^{({\cal B})}_a =  0 = \int_{Y_4} v \wedge [({\bf 3}, {\bf 2})] 
\end{split}
\end{align}
for any vertical divisor $D^{({\cal B})}_{a,b}$ and any $v \in V$. Then, for any particular flux solution $p = \sum \mu_k \, t_k$ satisfying (\ref{eq:transversality_condition}), (\ref{eq:gauge_symmetry_condition}) and (\ref{eq:no_stueckelberg}), with $\int_{Y_4} p \wedge [({\bf 3}, {\bf 2})] = 3$, the affine space $p + V$ clearly contains all fluxes giving rise to a spectrum with three generations of left-handed quarks. We choose the solution for $p$ to be as `short' as possible, i.e.~with smallest possible coefficients $\mu_k$. These coefficients are not necessarily integer due to the condition $\int_{Y_4} p \wedge [({\bf 3}, {\bf 2})] = 3$. Note that $p$ determined in this way is in general not properly quantised; this issue requires some further checks, see below.

For our scan, we determine a basis $\{b_i\}$ of $V$, s.t.~$G_4 = p + \sum_i \lambda_i \, b_i$, and then vary the $\lambda_i$ discretely over a finite range. Due to computational limitations, we have to restrict the range to be a subset of $[-10, 10]$, with the number of independent $\lambda_i$ ranging between 3 and 7, depending on the base and fibration data.
Because of the discrete increments, we need the lattice spanned by $\{b_i\}$ to be not too coarse. Furthermore, the basis vectors should have roughly equal `length', so that, by varying all $\lambda_i$ over the same range, we cover a `sphere' in $V$, i.e.~extending equally into all independent directions of the flux configuration space. This is accommodated by the following strategy:
\begin{itemize}
	\item The conditions (\ref{eq:subspace_V}) can be rearranged into a matrix whose $k$-th column is defined by the intersection numbers (\ref{eq:subspace_V}) with $v$ replaced by the basis vector $t_k$ of $H^{(2,2)}_\text{vert}(Y_4)$. The kernel of this matrix is $V \subset H^{(2,2)}_\text{vert}(Y_4)$, written in the basis $\{t_k\}$.
	\item Using \texttt{Sage}, we compute this kernel over $\mathbb{Z}$, i.e.~the resulting basis vectors $\{\tilde{b}_i\}$ are $\mathbb{Z}$-linear combinations of $\{t_k\}$.
	\item Finally we apply the Lenstra--Lenstra--Lov\'{a}sz (LLL) algorithm -- which is conveniently implemented in \texttt{Sage} -- to this set, yielding the basis $\{b_i\}$. The scan will then vary the coefficients $\lambda_i$ over the interval $[-10,10]$ in increments of 1.
\end{itemize}
The LLL algorithm computes a `short', `nearly' orthogonal lattice basis of the input lattice generated by $\{\tilde{b}_i\}$. 
Here, `orthogonality' is with respect to the bilinear form $t_i \cdot t_j \coloneqq \delta_{ij}$, which clearly is not the metric on $H^{(2,2)}$ induced by the intersection product, and therefore is irrelevant to us.
However, the attribute `short' -- which a priori is also with respect to the wrong metric -- is helpful to us, because the resulting flux basis $\{b_i\}$ is expressed with the smallest possible integer coefficients in terms of the $t_i$ (in practise mostly 0's and 1's). In our models, the intersection numbers $\int_{Y_4} t_i\,t_j$ are all of order 1 to 10, so arguably the $b_i$ are (up to factors of order 1) of the same length with respect to the intersection product.

With this basis, we find that when we vary different $\lambda_i$ with equal step-sizes, also the values of the chiral indices and D3-tadpole change in roughly equal increments.
We found in all our examples that the LLL-reduced basis $\{b_i\}$ is much more advantageous in this respect than the basis $\{\tilde{b}_i\}$, which is obtained by Gauss elimination. 
Having established the basis flux vectors, we compute for each set $\{\lambda_i\}$ the chiral indices by integrating the flux $p+\sum_i \lambda_i \, b_i$ over the matter surfaces listed in table \ref{tab:matter_surface_classes}.

We observe here that if we were to perform a similar search with the flux basis \eqref{eq:general_flux_basis}, then the condition of vanishing D-term for $\omega_Y$ would already introduce fractional coefficients. Furthermore, if we vary the coefficients in the basis \eqref{eq:general_flux_basis} in integer (or even half-integer) increments, the values for $\chi({\cal R})$ will change by much larger step-sizes  compared to the basis $\{b_i\}$. This makes the latter more practical for a scan. 

\subsubsection*{A note on the quantisation condition}

Let us comment briefly on the quantisation condition, $G_4 + c_2(Y_4)/2 \in H^{4}(Y_4, \mathbb{Z})$. Traditionally, it is a hard problem to systematically solve this condition for explicit geometries \cite{Collinucci:2010gz,Collinucci:2012as}.
We make no attempt of doing so within the scope of this work. In particular, our search algorithm presented above works with rational cohomology classes. Therefore a proper quantisation is not guaranteed.
Instead, we follow the usual method of performing a few sanity checks for each individual flux vector the algorithm produces. Specifically we check if 
\begin{align}
&\chi({\cal R}) \in \mathbb Z  \quad  \forall \,  {\cal R}, \qquad \int_{Y_4} \, \left( G_4 + \frac{c_2(Y_4)}{2} \right) \wedge D_i \wedge D_j \in \mathbb Z \, , \notag \\
& n_3 = \frac{1}{24}\chi(Y_4) - \frac{1}{2} \int_{Y_4} G_4 \wedge G_4 \in \mathbb Z \, , \notag
\end{align}
where the second condition is evaluated for any two toric divisors $D_{i,j}$.
Clearly, these are necessary conditions to be satisfied by a suitably quantised $G_4$-flux.

As remarked at the beginning of this section, we find that fluxes constructed with the basis $\{b_i\}$ in the above fashion are much more likely to be properly quantised than a flux constructed as a linear combination of the basis \eqref{eq:general_flux_basis}, whose fractional coefficients are determined by fixing certain chiral indices.

\subsection{Summary of Search Procedure}\label{sec:summary_search}

Here we give a short summary of the scope of the search and comment on the generic chiral spectrum.

In general, we found that apart from the actual scan over the parameter space of the $\lambda_i$'s, the most computation time consuming procedure is performing the triangulations of the toric polytopes defining $X_5$. E.g.~for the base choice ${\cal B} = {\rm Bl}_2 \mathbb{P}^3$, the triangulation of all 59 polytopes with \texttt{Topcom} took roughly 4 months.\footnote{The computation was carried out with an Intel E6700 (3.2GHz) dual-core CPU and 4GB RAM.}
For this reason, we restricted ourselves to the three simplest toric bases. To perform a broader scan more efficiently, one would certainly need to find a faster algorithm for triangulations of toric polytopes, perhaps similar to the strategy of \cite{Long:2014fba} developed for threefolds.

The simplest base we considered is the standard choice ${\cal B} = \mathbb{P}^3$, with usual homogeneous coordinates $[z_0, z_1, z_2, z_3]$. Up to coordinate re-definition this allows for one single choice $(w_2, w_3) = (z_0, z_1)$ for the coordinates of the non-abelian divisors. There are then 16 consistent fibrations, i.e.~pairs of classes $(\alpha, \beta)$ entering (\ref{eq:sections_classes}). For each of these 16 fibrations the number of basis vectors $b_i$ is between 3 and 5. Out of the 16 different fibrations, only one produced properly quantised fluxes with `reasonable' chiral indices (in the range $|\chi| < 10$) within our search process.

Next, the base ${\cal B}= {\rm Bl}_1 \mathbb{P}^3$ is obtained by blowing-up $\mathbb{P}^3$ in a point.  The blow-up coordinate $x$ and associated divisor class $X$ corresponds to the ray $(0,0,0,-1)$ in the toric description. There are several inequivalent choices for the coordinates $w_{2,3}$. We have analysed the two possibilities $(w_2, w_3) = (z_0, x)$ and $(w_2, w_3) = (x, z_0)$, or in terms of divisor classes, $(W_2, W_3) = (H, X)$ and $(W_2, W_3) = (X, H)$, with $H$ the hyperplane class of $\mathbb{P}^3$. The first choice gives rise to 36 different fibrations. Out of these we find none with fluxes leading to `reasonable' chiralities (in the range $|\chi| < 10$).
The second choice $(W_2, W_3) = (X, H)$ allows for 40 different fibrations, amongst which there are three with `reasonable' flux configurations.
These fibrations have 5 or 6 independent basis vectors $b_i$.

We have also attempted to extend our search algorithm to ${\cal B} = {\rm Bl}_2 \mathbb{P}^3$ with blowup coordinates $x$ and $y$ corresponding to the rays $(0,0,0,-1)$ and $(0,0,-1,0)$. With the choice $(w_2, w_3) = (x,y)$, only one of the two inequivalent triangulations of the polytope for ${\cal B}$ is phenomenologically interesting, namely the one for which $x\,y$ is not in the SR-ideal. 
As we have mentioned earlier, the reason is of course that we insist on the presence of the bifundamental states $({\bf 3}, {\bf 2})$, which are localised at the intersection. It turns out, however, that for many choices of $(\alpha, \beta)$ giving rise to effective classes \eqref{eq:sections_classes}, the resulting space $X_5$ does not exhibit a compatible fibration structure over $\cal B$ (with the chosen intersection property of $\{x\}$ and $\{y\}$). In addition, all these cases lead to dimension-one singularities in $X_5$, so they would generically induce point-like singularities on $Y_4$. Out of the 59 possible choices for $(\alpha, \beta)$, only 30 have compatible fibrations. Of these, none has a properly quantised flux solution leading to chiral indices smaller than 10.

All of the `reasonable' spectra we found do not reproduce the Standard Model exactly. They all have chiral exotics, which can potentially give rise to interesting Beyond-the-Standard-Model physics. However, in most cases, the excess is still too large to comfortably relate them with the Standard Model. In the following we will discuss one example which is closest to the MSSM. The remaining `reasonable' models are listed in appendix \ref{app:all_chiral models}.

\subsection{An almost Standard-Model-like Example}

The class of fibrations over ${\cal B} = {\rm Bl}_1 \mathbb{P}^3$ with $\overline{\cal K} = 4\,H+2\,X$ is parametrised by the two divisors 
\begin{align}
\alpha = \alpha_H \, H + \alpha_X \, X, \qquad \quad \beta = \beta_H \, H + \beta_X \, X,
\end{align}
where $X$ and $H$ denote the two independent divisors of ${\cal B}$. 
With only one possible triangulation of the toric polytope, $\cal B$ has the following independent intersection numbers:
\begin{align} \label{Bl1intnumb}
	\int_{\cal B} X^3 = 1 \, , \quad \int_{\cal B} X^2\wedge H = -1 \, , \quad \int_{\cal B} X\wedge H^2 = 1 \, , \quad \int_{\cal B} H^3 =0 \, .
\end{align}
With the choice $(w_2, w_3) = (x, z_0)$ corresponding to $(W_2, W_3) = (X, H)$, the ambient space $X_5$ can be described by the following polytope:
\begin{align}\label{tab:toric_data_Bl1P3_W2=X}
	\begin{array}{c|c|c|c|c|c|c|c|c|c|c|c|c}
		\su & \sv & \sw & s_0 & s_1 & e_1 & f_1 & f_2 & e_0 & f_0 & z_1 & z_2 & z_3	 \\ \hline
		-1 & 0 & 1 & -1 & 0 & 1 & 0 & 1 & 0 & 0 & \alpha_X - \alpha_H & 0 & -\alpha_X \\
		1 & -1 & 0 & 0 & 1 & 0 & 1 & 0 & 0 & 0 & \beta_H - \beta_X & 0 & \beta_X \\
		0 & 0 & 0 & 0 & 0 & 0 & -1 & -1 & 0 & -1 & 1 & 0 & 0 \\
		0 & 0 & 0 & 0 & 0 & 0 & -1 & -1 & 0 & -1 & 0 & 1 & 0 \\
		0 & 0 & 0 & 0 & 0 & -1 & -1 & -1 & -1 & -1 & 0 & 0 & 1
	\end{array}
\end{align}
As explained before, the coefficients $\alpha_{(\cdot)}, \beta_{(\cdot)}$ must be chosen such that the classes (\ref{eq:sections_classes}) are effective, i.e.~their expansion in $X$ and $H$ must have positive coefficients. There are 40 tuples $(\alpha_H, \alpha_X, \beta_H, \beta_X)$ satisfying this condition.

Within our scan, the flux configuration coming closest to the Standard-Model spectrum is based on the fibration defined by $\alpha = 3H + X$, $\beta = H + X$. The Euler number of the elliptic fourfold inside $X_5$ is $\chi(Y_4) =1794$.
The flux configuration of interest is furthermore defined for the hypercharge identification $U(1)_Y = U(1)_1$, and takes the form
\begin{align}\label{eq:vanilla_flux}
\begin{split}
	G_4  = \,& \frac{1}{2} \, ( E_1 \wedge (2\,H - 3\,F_2 - S_1) + X \wedge( F_2 - F_1 + S_1)) \\
	        = \, & -\frac{1}{6} G^{z_2}_4 - \frac{1}{12} G^{z_4}_4 \, .
\end{split}
\end{align}
In the second line we have identified this flux with the specialisation of general fluxes \eqref{eq:general_flux_basis} we derived in the generic setting to this particular fibration.
Explicit checks confirm that this flux satisfies all necessary conditions for being appropriately quantised: the intersection numbers $\int_{Y_4} (G_4 + c_2(Y_4)/2 ) \wedge D_i \wedge D_j$ are all integer, all the chiral indices are integer (see below), and the number of D3-branes required to cancel the D3-tadpole is
\begin{align}
n_3 = \frac{1}{24} \chi(Y_4) - \frac{1}{2} \int_{Y_4} G_4^2 = 72.
\end{align}
It turns out the flux induces a vanishing D-term not only for $U(1)_Y$, as required, but in fact for $U(1)_1$ and $U(1)_2$ individually,
\begin{align}
\int_{Y_4} G_4 \wedge \omega_1 \wedge D_a^{{\cal B}} = \int_{Y_4} G_4 \wedge \omega_2 \wedge D_a^{{\cal B}} = 0 \qquad \quad \forall D_a^{{\cal B}} \in H^{1,1}({\cal B}).
\end{align}
In particular, the flux (\ref{eq:vanilla_flux}) is not a $U(1)_i$ gauge flux.
Therefore, neither of the abelian gauge factors acquires a St\"uckelberg mass and  the gauge symmetry 
is $SU(3) \times SU(2) \times U(1)_Y \times U(1)_2$, with an extra massless abelian gauge group factor compared to the Standard Model.

The induced chiral spectrum is summarised in table \ref{eq:vanilla_spectrum} and follows directly from the expressions in tables \ref{tab:explicit_chiralities} and \ref{tab:chiralities_missing_singlets} with the help of the base intersection numbers (\ref{Bl1intnumb}).
As one can see, the largest (absolute value of) chirality is 4, which is among the smallest values we have been able to find within our search process; in particular this means that there are necessarily chiral exotics beyond the MSSM spectrum.
{\renewcommand{\arraystretch}{1.25}
\begin{table}[ht]
\centering
	\begin{tabular}{c||c|c|c|c|c|c|c|c}
			$\cal R$ & $ {\bf 2}_1$ & ${\bf 2}_2$ & ${\bf 2}_3$ & ${\bf 3}_1$ & ${\bf 3}_2$ & ${\bf 3}_3$ & ${\bf 3}_4$ & ${\bf 3}_5$ 
			\\
			$(q_1, q_2)$ & $\left( \frac{1}{2}, -1 \right)$ & $\left( \frac{1}{2}, 1\right)$ & $\left( \frac{1}{2}, 0 \right)$ & $ \left( \frac{2}{3}, -\frac{1}{3} \right)$ & $\left( -\frac{1}{3}, -\frac{4}{3} \right)$ & $\left( -\frac{1}{3}, \frac{2}{3} \right)$ & $\left( \frac{2}{3}, \frac{2}{3} \right)$ & $\left( -\frac{1}{3}, -\frac{1}{3} \right)$
			\\ \hline 
			$\chi$ & $-2$ & $1$ & $-2$ & $-2$ & $0$ & $1$ & $-1$ & $-4$    \\ \hline \hline \rule{0pt}{3ex} 
			${\cal R}$ & $({\bf 3}, {\bf 2})$ &  $ {\bf 1}^{(1)} $& ${\bf 1}^{(2)}$ & ${\bf 1}^{(3)}$ & ${\bf 1}^{(4)}$ & ${\bf 1}^{(5)}$ & ${\bf 1}^{(6)}$ 
			\\
			$(q_1, q_2)$ & $\left( \frac{1}{6}, -\frac{1}{3} \right)$ & $(1,-1)$ & $(1,0)$ & $(1,2)$ & $(1,1)$ & $(0,2)$ & $(0,1)$ 
			\\ \cline{1-8} 
			$\chi$ & $3$ & $2$ & $1$ & $0$ & $0$ & $0$ & $-4$
	\end{tabular}
	\caption{The chiral spectrum induced by the flux \eqref{eq:vanilla_flux}. For completeness we have included the $U(1)_1 \times U(1)_2$ charges $(q_1, q_2)$ of the states.}\label{eq:vanilla_spectrum}
\end{table}}

To actually make contact with particle physics, we invoke our classification of possible Standard Model matchings from appendix D in \cite{Lin:2014qga}. Specifically, we consider the possibility no.~7 in table D.1 of \cite{Lin:2014qga}.
 This leads to the identifications listed in table \ref{tab:vanilla_spectrum_matchings}.
 The specification `heavy' or `light' refers to whether or not a perturbative Yukawa coupling of order one with the Higgs field is generated. Indeed, as analysed in more detail in \cite{Lin:2014qga}, all Yukawa couplings which are allowed by the non-abelian and abelian gauge symmetry are realised geometrically 
 at points on ${\cal B}$ where the associated matter curves intersect.
 \begin{table}[ht]
\begin{center}
	\begin{tabular}{c||c|c|c|c|c|c|c|c}
		$\cal R$ & $\overline{\bf 2}_1$ & ${\bf 2}_2$ & $\overline{\bf 2}_3$ & $\overline{\bf 3}_1$ & $\overline{\bf 3}_2$ & $\overline{\bf 3}_3$ & $\overline{\bf 3}_4$ & $\overline{\bf 3}_5$  \\ %
		\hline \rule{0pt}{2.5ex}
		$\chi$ & $2$ & $1$ & $2$ & $2$ & $0$ & $-1$ & $1$ & $4$  \\ %
		\hline \rule{0pt}{2.5ex}
		SM & \multirow{2}{*}{$L$} & \multirow{2}{*}{$H_u$} & \multirow{2}{*}{$L + H_d$} & light & light & light & heavy & heavy \\ %
		states  & & & & $u_R^c$ & $d_R^c$ & $d_R^c$ & $u_R^c$ & $d_R^c$ \\[.5ex] %
		\hline \hline \rule{0pt}{3ex}
		$\cal R$ & $({\bf 3}, {\bf 2})$  & ${\bf 1}^{(1)}$ & ${\bf 1}^{(2)}$ & ${\bf 1}^{(3)}$ & ${\bf 1}^{(4)}$ & ${\bf 1}^{(5)}$ 	& \multicolumn{2}{c}{$\overline{\bf 1}^{(6)}$}\\ 
		\cline{1-9} \rule{0pt}{2.5ex}
		$\chi$ & $3$ & $2$ & $1$ & $0$ & $0$ & $0$ 
		& \multicolumn{2}{c}{$4$}\\
		\cline{1-9} \rule{0pt}{2.5ex}
		SM & \multirow{2}{*}{$Q$} &  heavy & heavy & \multirow{2}{*}{$-$} & heavy & \multirow{2}{*}{$-$} 
		& \multicolumn{2}{c}{heavy $\nu_R^c$,}\\
		states  & \, & $e_R^c$ & $e_R^c$ & &$e_R^c$ & 
		& \multicolumn{2}{c}{$\mu$-term}
	\end{tabular}
\end{center}
\caption{Possible matching of the chiral spectrum obtained from the flux (\ref{eq:vanilla_flux}) with the (N)MSSM spectrum.}
\label{tab:vanilla_spectrum_matchings}
\end{table}

The exotics which do not fit into the MSSM are a pair of triplets residing on the curves ${\bf 3}_3$ and ${\bf 3}_5$, as well as the singlets on ${\bf 1}^{(6)}$. If indeed the chirality 2 for $\overline{\bf 2}_3$ is distributed as 1 for $H_d$ and 1 for the leptons $L$, then the Higgs $(H_u, \overline{H_d}) $ come as a vector-like pair. Likewise, the excess of chiral triplets can be grouped into a vector-like pair $ (\overline{\bf 3}_3)^c + \overline{\bf 3}_5 $ charged like the Standard-Model down-quarks. In light of recent events at the LHC, these exotics could possibly be of interest (e.g.~in the spirit of \cite{Cvetic:2015vit,Cvetic:2016omj,Palti:2016kew}), but we will not attempt any detailed phenomenological discussion in this direction.
Irrespective of the question of exotics, the model must be considered in the context of intermediate or high scale supersymmetry breaking because the charge assignments and resulting Yukawa couplings give rise to dimension-four proton decay operators which would be incompatible with a TeV supersymmetry scale. 
This happens despite the appearance of the extra $U(1)_2$ selection rule. The complete list of such operators can be found in table D.1 of \cite{Lin:2014qga}.

Finally, note that we have only computed the chiral spectrum. On top of this, extra vector-like pairs of massless matter localised over a single curve may exist. Their computation, e.g.~along the lines of \cite{Bies:2014sra}, is considerably more involved and beyond the scope of this work.

To arrive at the precise Standard-Model spectrum, and to remove the extra massless $U(1)_2$ from the spectrum, one can imagine Higgsing the latter with a vector-like pair of massless singlets ${\bf 1}^{(6)} + \overline{\bf 1}^{(6)}$ in a D-flat manner,
\begin{align} \label{6Higgsing}
\langle {\bf 1}^{(6)} \rangle = \langle {\bf \overline 1}^{(6)} \rangle \neq 0.
\end{align}
This of course assumes that at least one such vector-like pair is available.
The recombination singlets couple to the massless matter multiplets as follows \cite{Lin:2014qga}:
\begin{align}\label{6cuplings}
\begin{split}
{\cal L} \supset&  \,  {\bf 1}^{(6)} \,  {\bf 3}_2 \, {\bf \overline 3}_5  +  {\bf 1}^{(6)} \,  {\bf 3}_1 \, {\bf \overline 3}_4 +   {\bf \overline 1}^{(6)} \,  {\bf 3}_3 \, {\bf \overline 3}_5 + {\bf 1}^{(6)} \,  {\bf 2}_1 \, {\bf \overline 2}_3 + {\bf \overline 1}^{(6)} \,  {\bf 2}_2 \, {\bf \overline 2}_3 +   \\
&  \,  {\bf 1}^{(6)} \,  {\bf 1}^{(2)} \, {\bf \overline 1}^{(4)} + {\bf 1}^{(6)} \,  {\bf 1}^{(1)} \, {\bf \overline 1}^{(2)} + {\bf 1}^{(6)} \,  {\bf 1}^{(4)} \, {\bf \overline 1}^{(3)} + {\bf 1}^{(6)} \,  {\bf 1}^{(6)} \, {\bf \overline 1}^{(5)} + c.c.
\end{split}
\end{align}
Note that the last term would induce an F-term for ${\bf \overline 1}^{(5)}$ in the background (\ref{6Higgsing}).
It therefore comes as a relief that 
the chiral index associated with the singlets ${\bf 1}^{(5)} + {\bf \overline 1}^{(5)}$ is indeed vanishing, see table 
 \ref{eq:vanilla_spectrum}. Absence of an F-term obstruction to the recombination then requires that in addition no vector-like pair ${\bf 1}^{(5)} + {\bf \overline 1}^{(5)}$ of massless such singlets exist.
If this condition is satisfied, the Higgsing (\ref{6Higgsing})
 recombines the curves ${\bf 3}_1$ and ${\bf 3}_4$, the curves ${\bf 3}_3$ and ${\bf 3}_5$, furthermore all the ${\bf 2}_i$ curves as well as the singlet curves ${\bf 1}^{(1)}$, ${\bf 1}^{(2)}$ and ${\bf 1}^{(3)}$, in agreement with the couplings (\ref{6cuplings}).
 This  leads to the following spectrum:
\begin{center}
	\begin{tabular}{c||c|c|c|c}
		state & $({\bf 3}, {\bf 2})$ & $\overline{\bf 3}_1 + \overline{\bf 3}_4$  & $\overline{\bf 3}_3 + \overline{\bf 3}_5$ & $\overline{\bf 2}_1 + \overline{\bf 2}_2 + \overline{\bf 2}_3$ \\
		\hline \rule{0pt}{2ex}
		$\chi$ & $3$ & $3$ & $3$ & $3$ \\
		\hline \rule{0pt}{2ex}
		SM & \multirow{2}{*}{$Q$} & \multirow{2}{*}{$u_R^c$} & \multirow{2}{*}{$d_R^c$} & \multirow{2}{*}{$L + H_u + H_d$} \\
		states & & & &\\
		\hline \hline \rule{0pt}{3ex}
		state & ${\bf 1}^{(1)} + {\bf 1}^{(2)} + {\bf 1}^{(4)}$ & ${\bf 1}^{(5)}$  & ${\bf 1}^{(6)}$  & $\overline{\bf 1}^{(6)}$\\ 
		\hline \rule{0pt}{2ex}
		$\chi$ & $3$ & $0$ & $-$ & $-$\\
		\hline \rule{0pt}{2ex}
		SM & \multirow{2}{*}{$e_R^c$} & \multirow{2}{*}{$-$} & \multirow{2}{*}{vev} & \multirow{2}{*}{vev} \\
		states  & & & &
	\end{tabular}	
\end{center}
Note that now the Higgs-doublet is localised on the same curve, similar to the 3-chiral generation MSSM realised in \cite{Cvetic:2015txa}.
Phenomenological viability therefore requires one extra massless vector-like pair of associated states after the recombination.



\section{Conclusions and Outlook } \label{sec_concl}

In this work we have classified the vertical gauge fluxes for one of the five inequivalent F-theory fibrations introduced in \cite{Lin:2014qga} which gives rise to the Standard Model group plus an extra abelian gauge group factor. Our analysis of the vertical cohomology ring and the computation of the chiral indices of the charged matter have been performed in a manner independent of a choice of base space of the fibration. The obtained expressions, in particular the results in tables \ref{tab:explicit_chiralities} and \ref{tab:chiralities_missing_singlets} for the chiral indices, are {\it pr\^{e}t-\`a-porter}: They are immediately applicable to any base ${\cal B}$ compatible with the fibration structure. 
As a first such application we have searched for three-generation models on the base spaces ${\cal B} \in \{ \mathbb P^3, {\rm Bl}_1\mathbb P^3, {\rm Bl}_2\mathbb P^3$ \}. The second base supports a fully consistent flux configuration giving rise, at the chiral level, to the Standard Model spectrum plus an extra triplet pair. The latter can be lifted, under certain assumptions about the vector-like spectrum, by a recombination process. 
Clearly it would be desirable to extend the analysis to the remaining four types of fibrations introduced in \cite{Lin:2014qga}, and furthermore perform a broader search based on an improved triangulation algorithm that allows for a more time efficient scan with other base spaces.

Both from a conceptual perspective and as motivated by phenomenological considerations an important step forward is to go beyond the computation of merely the chiral index of charged matter in F-theory. A formalism for how to approach this important task has been presented in \cite{Bies:2014sra}, based on a finer parametrisation of the flux data than merely in terms of the gauge fluxes. It will be interesting to apply this philosophy to the fibrations studied in this article: In a first step one will have to provide a refined description of the vertical gauge data in terms of the Chow ring on $Y_4$. The second task is to perform the intersection theoretic pairing with the matter surfaces in such a way that we can extract the line bundles on the matter curves on ${\cal B}$ whose cohomology counts the exact massless spectrum.

Another off-spring of the general parameterisation of the matter surfaces obtained in this work is a new and very general approach to understanding the constraints which the cancellation of gauge anomalies imposes on the geometry of elliptic fourfolds: 
We have been able to verify, for the concrete fibrations under consideration, that the pure and mixed non-abelian anomalies are automatically cancelled for any consistent gauge flux. This is a general consequence of the structure of the matter surfaces rather than of the explicit form of $G_4$. More precisely, the relevant combinations of matter surfaces entering the anomalies and the Green--Schwarz counterterms \cite{Cvetic:2012xn} are observed to be automatically annihilated by any vertical flux satisfying the transversality conditions (\ref{eq:transversality_condition}) and the condition (\ref{eq:gauge_symmetry_condition}) of unbroken non-abelian gauge symmetry in the F-theory limit. 
This observation calls for a general proof in terms of  properties of the matter surfaces for any consistent fibration. This would help us to establish a deeper geometric interpretation of anomaly cancellation directly in terms of the middle (co-)homology of elliptic fourfolds, complementary to the analysis of \cite{Grimm:2015zea}. Such an understanding would be the four-dimensional counterpart to the geometrisation of the anomaly cancellation conditions for F-theory on Calabi--Yau  threefolds as put forward in \cite{Grassi:2000we,Grassi:2011hq}.

\subsection*{Acknowledgements}
We are indebted to Christoph Mayrhofer and Oskar Till for many important discussions. 
The work of L.L.~was supported by a scholarship of the German National Academic Foundation. 
This research was supported in part by DFG under TR 33 `The Dark Universe'.


\newpage

\begin{appendix}

\section{Vertical Cohomology of toric Hypersurfaces}\label{sec:calculating_flux_basis}

In this appendix we describe the vertical cohomology ring of a toric fibration over a generic base by an appropriate quotient ring.

A vertical cohomology form $\omega$ of degree $(k,k)$ is a linear combination of wedge products of $k$ $(1,1)$-forms Poincar\'{e}-dual to divisors $\text{PD}(D^{(Y_4)}_i) \equiv D^{(Y_4)}_i \in H^{(1,1)}(Y_4)$. Recall from section \ref{sec:vertical_fluxes} that in our setup the elliptic fibration $Y_4 \rightarrow {\cal B}$ is the zero locus of a polynomial $P_T$ inside an ambient space $X_5$, which is a fibration of a toric fibre ambient space over the same base ${\cal B}$. In our constructions, all divisors on the hypersurface $Y_4 = \{P_T\}$ are restrictions of divisors from the ambient space $X_5$, i.e.~$D^{(Y_4)}_i \cong D_i \cap \{P_T\}$ for some $D_i \in H^{(1,1)}(X_5)$.\footnote{We use `$\cong$' to take note of the fact that $D^{(Y_4)}_i$ is a priori defined on $Y_4$, however, by the embedding of $Y_4 \xhookrightarrow{i} X_5$ one can identify $D^{(Y_4)}_i \equiv i_*(D_i) = D_i \cap \{P_T\}$ in the ambient space homology.}
The product $\bigwedge_i D^{(Y_4)}_i = \omega$ on $Y_4$ corresponds in homology to the intersection product $\bigcap_i D_i \cap \{P_T\} \equiv \tilde{\omega} \cap \{P_T\} \subset X_5$. 
On $Y_4$ the wedge product of $\omega$ with another vertical form $\eta$ is then Poincar\'{e}-dual to $\tilde\omega \cap \tilde\eta \cap \{P_T\} \in H_\text{vert}^{(\cdot, \cdot)}(X_5)$, with $\tilde{\eta} \cap \{P_T\} \cong \eta$. It is not hard to see that the result does not depend on the choice of the representatives $\tilde\omega$ and $\tilde\eta$. In fact, the ambiguity precisely stems from the linear equivalence relations and intersection properties among divisors of the ambient space. In the toric case, this information is contained in the linear equivalence ideal (LIN) and Stanley--Reisner ideal (SRI), respectively. For simplicity we will stick to these terms for fibrations over any base (we will define what we mean by the SRI in this case), even if the resulting ambient space is not toric.

For our calculations it is useful to think of a vertical $(k,k)$-form $\tilde\omega$ on the ambient space as a polynomial expression in a set of divisors $D_i$ with coefficients in $\mathbb{Q}$, i.e.~$\tilde\omega \in \mathbb{Q}[D_i]$. 
In this representation a wedge product of forms is just given by polynomial multiplication modulo the linear relations (LIN) and intersection properties (SRI). The set LIN contains all linear combinations of divisors which are zero in homology. Now the sum of two such linear combinations as well as any $(k,k)$-form wedged with such a linear combination is still zero, therefore LIN is an actual ideal of $\mathbb{Q}[D_i]$. 
Following the toric geometry setup, we define the ideal SRI for any fibration to be generated by all formal products of divisors which are zero in (co-)homology. With this, two polynomials in $\mathbb{Q}[D_i]$ represent the same class in (co-)homology if they differ by an element of the form $s+l$ with $s \in {\rm SRI}$ and $l \in {\rm LIN}$. Therefore, the multiplicative structure of the cohomology ring is fully encoded in the quotient ring
\begin{align}\label{eq:cohom_ambient_space_general}
	\bigoplus_k H^{(k, k)}_\text{\tiny vert} (X_5, \mathbb{Q}) \cong \frac{\mathbb{Q}[D_i]}{{\rm SRI} + {\rm LIN}} \, ,
\end{align}
where the denominator is the ideal generated by both SRI and LIN (and corresponds to the sum of the ideals). Note that the grading of the cohomology ring is simply given by the natural grading of polynomials by their degree.

From the above discussion, it also immediately follows that we can represent the vertical cohomology of the hypersurface as
\begin{align}\label{eq:cohom_ring_fourfold}
	H^{(k,k)}_\text{\tiny vert} (Y_4, \mathbb{Q}) \cong \frac{ \mathbb{Q} [D_i]^{(k)} \wedge [P_T] }{\text{SRI}+\text{LIN}} \subset H^{(k+1, k+1)}_\text{vert} (X_5, \mathbb{Q}) \, ,
\end{align}
where $\mathbb{Q} [D_i]^{(k)}$ denotes polynomials in $\mathbb{Q} [D_i]$ with only degree $k$ monomials.
This formalism, which also underlies the analysis of \cite{Lin:2015qsa}, has previously been applied in \cite{Krause:2012yh} to classify the vertical cohomology groups for the $SU(n)$ Tate models with $n \leq 5$ over general bases and in \cite{Grimm:2011fx,Braun:2013yti,Cvetic:2013uta,Cvetic:2015txa}  for various fibrations over concrete base manifolds.

This now effectively allows us to carry out all relevant computations in $H^{(k,k)}_\text{\tiny vert} (Y_4, \mathbb{Q})$ in the ambient space cohomology.
For explicit computations we implement the quotient ring structure (\ref{eq:cohom_ring_fourfold}) into \texttt{Singular} \cite{singular}, which is designed for calculations within polynomial rings. The results in sections \ref{sec:gauge_anomalies} and \ref{sec:witten_anomaly} are all computed in this setup. 
Furthermore, \texttt{Singular} can also readily compute the minimal generating set of an ideal, expressed as monomials in the ring variables $D_i$. Applied to the ideal $\mathbb{Q}[D_i]^{(2)} \wedge [P_T] / (\text{SRI}+\text{LIN})$, this in particular gives a vector space basis $\{t_i = D_{a_i} \wedge D_{b_i} \}$ of $H^{(2,2)}_\text{vert}(Y_4, \mathbb{Q})$, which is the starting point of determining a basis of $G_4$-fluxes.

\subsubsection*{Fibrations with generic Base}

By a fibration over a generic base we mean a setup in which different vertical divisors are treated as linearly independent, and in which intersection products on the base are always non-zero unless the codimension of the intersection exceeds the dimension of the base. To describe such geometries we mimic the vertical cohomology with a quotient ring of the form
\begin{align}\label{eq:cohomology_toric_ambient_space}
	\bigoplus_k H^{(k,k)}_{\text{vert}} (X_5, \mathbb{Q}) \cong \frac{ \mathbb{Q} [D^{(T)}_i , D^{({\cal B})}_j ] }{\text{SRI}^{(T)} + \text{SRI}^{({\cal B})} + \text{LIN}^{(T)}  } \, ,
\end{align}
where we split the set of divisors into those that come from the top ($D^{(T)}_i$) and the vertical divisors from the base ($D^{({\cal B})}$). Since the top we use to define the fibration fully specifies the fibre ambient space and parts of the fibration data, it relates fibral divisors linearly to one another and to certain vertical divisors, forming the ideal ${\rm LIN}^{(T)}$ (in the ${\rm I} \times {\rm A}$ model we studied, this is given by (\ref{eq:LIN-generators})). 
The genericness of ${\cal B}$ is implemented by assuming no further linear relations amongst the vertical divisors (i.e.~no contributions to the ideal LIN involving only base divisors). The ideal SRI in \eqref{eq:cohom_ambient_space_general} will have a part ${\rm SRI}^{(T)}$ coming from the Stanley--Reisner of the top (cf.~\eqref{eq:SR-ideal-IxA}), as well as a part ${\rm SRI}^{({\cal B})}$ encoding intersection properties of the base. 
Since any intersection with the allowed codimension is a priori non-zero for a generic base, we only have to ensure that any intersection product with more than `three legs on the base' vanishes, as it should for a fibration over a threefold base ${\cal B}$. This is realised if we define the ideal ${\rm SRI}^{({\cal B})}$ to be generated by
\begin{align*}
	D^{({\cal B})}_1 \wedge D^{({\cal B})}_2 \wedge \left\{
		\begin{array}{l}
			D^{({\cal B})}_3 \wedge D^{({\cal B})}_4\\
			D^{({\cal B})}_3 \wedge {\rm Ex}_i \\
			{\rm Ex}_i^{G_1} \wedge {\rm Ex}^{G_2}_j
		\end{array}
	\right\}
\end{align*}
for any vertical divisor $D_k^{({\cal B})}$ and any exceptional divisor ${\rm Ex}_i$. Here $G_1$ and $G_2$ refer to two independent gauge algebras realised on two different divisors on ${\cal B}$. 

Aside from the vertical divisors that have non-trivial linear relations with fibral divisors (in our models these are $\alpha, \beta, \overline{\cal K}, W_2$ and $W_3$, cf.~section \ref{sec_Anomalies}), any other vertical divisor will appear on equal footing for a generic base. This can be mimicked by introducing a further `dummy' vertical divisor $D$ as a formal variable of the polynomial ring (\ref{eq:cohomology_toric_ambient_space}) which is not related to any of the divisors $D^{(T)}$ by linear relations.
This dummy divisor comes in handy e.g.~in computations involving the $U(1)$-fluxes \eqref{eq:general_flux_basis} or the D-terms \eqref{eq:D-terms_generic_base}.
We stress that the resulting quotient ring is not truly a cohomology ring, e.g.~it does not satisfy Poincar\'{e}-duality, $\dim \, H_{\text{\tiny vert}}^{k,k} = \dim \, H_{\text{\tiny vert}}^{5-k,5-k}$. However it captures the essential features needed e.g.~to compute the general fluxes \eqref{eq:general_flux_basis}, or the homology classes of matter surfaces and the anomalies in section \ref{sec_Anomalies}.
 Furthermore, note that when one specifies a concrete base ${\cal B}$, all these `generic' relations remain true, but they may be completed by extra linear relations from the specific intersection structure on ${\cal B}$. All conclusions drawn from the `generic' relations remain valid for such specialisations.

It is worth noting that it is generally possible to reduce ~intersections of five divisors in the ambient space to a sum of intersection numbers of three divisors on the base if the fibre ambient space is fully specified. In our implementation of (\ref{eq:cohomology_toric_ambient_space}) into \texttt{Singular}, this is reflected as follows:
Any degree five polynomial $P^{(5)} \in H^{(5,5)}_\text{vert}(X_5)$ will be reduced into an expression of the form 
\begin{align}
(\sum D^{({\cal B})}_a \, D^{({\cal B})}_b \, D^{({\cal B})}_c ) \wedge ( \sum D_i^{(T)} \, D^{(T)}_j) \equiv \# \cdot \int_{\cal B} (\sum D^{({\cal B})}_a \, D^{({\cal B})}_b \, D^{({\cal B})}_c ) \, ,
\end{align}
 with a specific quadratic term $\sum D_i^{(T)} \, D^{(T)}_j$ which is the same for any polynomial $P^{(5)}$.\footnote{For this to be the case, the polynomial ring (\ref{eq:cohomology_toric_ambient_space}) has to be defined in \texttt{Singular} with the appropriate monomial ordering. We always used `degree reverse lexicographical ordering' (`dp'), in which case the variables for the divisors have to be put in such an order that the base divisors are listed \textit{after} the top divisors. The specific universal factor $ \sum D_i^{(T)} \, D^{(T)}_j$ may depend on the ordering of the fibral coordinates.}
To infer the numerical prefactor the quadratic term represents, one can simply reduce a universally known intersection number of the fibration: E.g.~in a model with a section $S_0$ one reduces the expression $S_0 \wedge [P_T] \wedge (\sum D^{({\cal B})}_a \, D^{({\cal B})}_b \, D^{({\cal B})}_c )$, which we know is equal to $1 \cdot \int_B (\sum D^{({\cal B})}_a \, D^{({\cal B})}_b \, D^{({\cal B})}_c )$. Analogously for a model with a bisection $U$, as in \cite{Lin:2015qsa}, we compare to $U \wedge [P_T] \wedge (\sum D^{({\cal B})}_a \, D^{({\cal B})}_b \, D^{({\cal B})}_c ) = 2 \cdot \int_B (\sum D^{({\cal B})}_a \, D^{({\cal B})}_b \, D^{({\cal B})}_c )$.

\subsubsection*{Comment on Fibrations with non-toric Divisors}

The above discussion applies to situations in which all divisors on the fibration $Y_4$ are inherited from divisors of the toric ambient space $X_5$.
This allowed us to use the intersection theory of $X_5$, where the linear equivalence (LIN) and intersection relations (SRI) are particularly easy to obtain.
More generally, however, some of the divisor classes on $Y_4$ may not be of this form. 
Such non-toric divisors  $D^{\rm nT}$ (e.g.~non-toric sections \cite{Mayrhofer:2012zy,Braun:2013nqa,Klevers:2014bqa, Cvetic:2015ioa}) are given in terms of vanishing loci of polynomials on $Y_4$ rather than the vanishing of toric ambient space coordinates. In practise, one can then analyse the intersection behaviour with another divisor $D$ by studying the set-theoretic intersection of $D^{\rm nT}$ with a \emph{generic} representative of $D$. 
With this knowledge the formalism developed in the previous section is readily extended:
Suppose that, in addition to all divisors $D^{(Y_4)}_i$ of the fibration, we also know the linear equivalence relations among the $D^{(Y_4)}_i$ (as the set ${\rm LIN}^{(Y_4)}$ defined directly on $Y_4$) and also which divisors do not intersect on $Y_4$ (${\rm SRI}^{(Y_4)}$), then the above considerations will obviously lead to the identification
\begin{align}
	\bigoplus_k H^{(k, k)}_\text{vert}(Y_4, \mathbb{Q}) \cong \frac{\mathbb{Q} [ D^{(Y_4)}_i ] }{ {\rm LIN}^{(Y_4)} + {\rm SRI}^{(Y_4)}} \, ,
\end{align}
without any reference to an ambient space.
It would be interesting to systematically explore fluxes  in such geometries in future work.


\section{Determining Homology Classes of Matter Surfaces}\label{app:homology_classes}

In order to calculate the chiral index (\ref{eq:calculate_chiral_index}) we need to determine the homology classes of the matter surfaces of the representations listed in table \ref{tab:spectrum_IxA}. To this end we first require an appropriate description of the matter surfaces in terms of vanishing loci ${\rm V}(I)$ of suitable prime ideals $I$ in the coordinate ring of the ambient space $X_5$. 
For this purpose we extend the `prime ideal technique' \cite{Cvetic:2013uta, Cvetic:2013jta, Lin:2014qga}, which before was introduced to determine the complicated singlet loci within the function ring of the base, to the full ambient space $X_5$. Concretely, this means that we now work with the polynomial ring generated by the coordinates in the fibre and the sections of the base, whose ideals define subvarieties of $X_5$.\footnote{Note that this is in no way related to the polynomial ring (\ref{eq:cohomology_toric_ambient_space}) representing the cohomology ring!} 
The methods of primary decomposition, saturation, and determining the dimension of ${\rm V}(I)$ for an ideal $I$ carry over directly.\footnote{The mathematical framework in which these computations are performed is known as the theory of Gr\"obner bases. For a short and hands-on description in an F-theory setup, see section 2.1 of \cite{Lin:2014qga}.} 
In \cite{Lin:2015qsa} we have already applied this method without much explanation in a different model to determine the homology classes there. In the following, we  review the method in a slightly more general setup which allows for future applications in various contexts. This material has already appeared in \cite{LingPoster} (see also \cite{Cvetic:2015txa}). 

We start with the observation that all matter charged under non-abelian gauge groups is localised over curves of the form $C = \{w\} \cap \{p\}$ in the base, where $[\{w\}]$ is the divisor supporting the gauge symmetry, and $p$ is a polynomial in the sections on ${\cal B}$ which themselves appear as coefficients in the hypersurface polynomial $P_T$ (cf.~table \ref{tab:spectrum_IxA}). On the fourfold, matter arises from the splitting of the fibres of certain exceptional divisors ${\rm Ex}_i$ over the curve $C$. In other words, the variety $\{{\rm ex}_i\} \cap \{p\} \cap \{P_T\} \subset X_5$ is reducible. 
In fact, the irreducible components are the matter surfaces of interest, which are  4-cycles, i.e.~codimension 3 in $X_5$. Their corresponding prime ideals can be found by decomposing the ideal $\langle {\rm ex}_i, p, P_T \rangle$ generated by the polynomials ${\rm ex}_i$, $p$ and $P_T$ within the coordinate ring. 
If $p$ is one of the sections (\ref{eq:sections_classes}), then $P_T|_{p={\rm ex}_i=0} = \Pi_k Q^{m_k}_k$ necessarily factors into irreducible polynomials $Q_k$, thus the resulting irreducible components are complete intersections with prime ideals of the form $\langle {\rm ex}_i, p, Q_k \rangle$ with multiplicity $m_k$. In this simple case the homology class of the irreducible component is just ${\rm Ex}_i \wedge [p] \wedge [Q_k]$.

However, if $p$ is a more complicated polynomial, then one cannot simply evaluate $P_T|_{p={\rm ex}_i=0}$ in the above fashion to factorise $P_T$. The usual procedure by solving $p =0$ for one of its variables and plugging the result into $P_T$ will generically introduce fractional or irrational expressions, which is no longer well-defined globally. Instead, when we perform the primary decomposition of $\langle {\rm ex}_i, p, P_T \rangle$ in this case, we will in general find associated irreducible components generated by more than three polynomials despite being codimension 3. The associated homology class is clearly no longer simply the product of the divisor classes of the individual generators. Instead, it will be the sum of several complete intersections.

To see this, assume that we have the codimension $d$ irreducible variety $\gamma = {\rm V}(I) \equiv {\rm V}( \langle f_1, ..., f_n \rangle)$ with $n>d$. Now consider the ideal $J$ generated by $d$ of the generators, w.l.o.g.~$J = \langle f_1, f_2, ..., f_d \rangle$. In general, $J$ will be reducible and decompose into some prime ideals $J^{(m)}$ with multiplicities $\mu^{(m)}$. Clearly, since $J \subset I$ (i.e.~${\rm V}(I) \subset {\rm V}(J)$), one of the ideals must be $I$ (with multiplicity $\mu$), say $J^{(0)} = I$. With a suitable choice of the $d$ generators, all the other components in the decomposition will be complete intersections of codimension $d$, i.e.~the prime ideal $J^{(m)}$ with $m \neq 0$ will have $d$ generators $f^{(m)}_1 , ..., f^{(m)}_d$. Since the class of a complete intersection is just the product of the classes of each generator, we have
\begin{align}
	\left[ {\rm V}(J) \right] &= \mu \left[ {\rm V}(I) \right] + \sum_{m\neq 0} \mu^{(m)} \left[ {\rm V}(J^{(m)}) \right] = \mu \, [ \gamma ] + \sum_{m \neq 0} \mu^{(m)} \bigwedge_{k=1}^d \left[ f^{(m)}_k \right] \notag \\
	\Longleftrightarrow \qquad [\gamma] &= \frac{1}{\mu} \left( \left[ {\rm V}(J) \right] - \sum_{m \neq 0} \mu^{(m)}  \left[ V \left(J^{(m)} \right) \right] \right)    = \frac{1}{\mu} \left( \bigwedge_{k=1}^d \left [f_k \right] - \sum_{m \neq 0} \mu^{(m)}  \bigwedge_{k=1}^d \left[ f^{(m)}_k \right] \right) \, .
	\label{eq:class_of_matter_surfaces}
\end{align}

In practise, if we have the prime ideal of a non-complete-intersection matter surface $\gamma$, we choose three of the generators and form a new ideal $J$, which we decompose into primary ideals. With a suitable choice the resulting prime ideals will all -- except for $I(\gamma)$ -- have three generators. To get their multiplicity, one has to compute the ideal saturation of $J$ with respect to the primes. We relied on \texttt{Singular} to perform the primary decomposition and to compute the ideal saturation to determine the irreducible components and their multiplicities (cf.~the next subsection for an explicit example).

Note that for the singlet states ${\bf 1}^{(i)}, i=2,4,6$, for which the curves $C^{(i)} = {\rm V}(I_{C^{(i)}})$ are not complete intersections, \texttt{Singular} is computationally unable to decompose the ideal $I_{C^{(i)}} + \langle P_T \rangle$ (which is the ideal describing the elliptic fibration restricted to the curve) into its irreducible components. Therefore we also cannot determine the homology classes of the corresponding matter surface by the above method.

\subsection[Matter Homology Classes in Model \texorpdfstring{${\rm I} \times {\rm A}$}{I x A} ]{Matter Homology Classes in Model {\boldmath{${\rm I} \times {\rm A}$}} }

We now exemplify the method outlined above to determine the homology classes of matter surfaces. For the representations ${\bf 2}^{\rm I}_1$, ${\bf 3}^{\rm A}_1$, ${\bf 3}^{\rm A}_2$, $({\bf 3}, {\bf 2})$, ${\bf 1}^{(1)}$, ${\bf 1}^{(3)}$ and ${\bf 1}^{(5)}$ the corresponding matter surfaces are complete intersections, thus their homology classes are clear: E.g.~a weight of ${\bf 2}^{\rm I}_1$ (in fact the highest weight) is given by $\{c_{2;0,1}\} \cap \{e_0\} \cap \{p\}$ where $p$ is an irreducible factor of the hypersurface equation restricted to $c_{2;0,1} = e_0 =0$; the homology class is thus $[c_{2;0,1}] \wedge E_0 \wedge [p]$. The exact form of $p$ can be found in table \ref{tab:matter_surface_classes}.

For the other representations, the matter surfaces are not complete intersection. As a simple example, let us look at ${\bf 3}^{\rm A}_3$ over the curve $\{w_3\} \cap \{b_{0;1,1}\,w_2\,c_{1;0,0} - b_1\,c_{2;0,1}\}$. From a local analysis (i.e.~solving the second equation defining the curve for one coefficient and plugging it into the hypersurface polynomial $P_T$), one finds that the weights arise from the splitting of the fibres of the resolution divisor $F_1$. The primary decomposition of the reducible surface $\{ f_1\} \cap \{ b_{0;1,1}\,w_2\,c_{1;0,0} - b_1\,c_{2;0,1} \} \cap \{P_T \}$ yields two prime components $I$ and $\bar{I}$ corresponding to weights of the fundamental and anti-fundamental representation. The component $\gamma_I$ defined by the prime ideal
\begin{align*}
	I = \langle \quad f_1, \quad b_{0;1,1}\,w_2\,c_{1;0,0} - b_1\,c_{2;0,1}, \quad c_{1;0,0}\, e_1 \, s_1\, w + c_{2;0,1} \, f_0 \, s_0 \, \sv , \quad b_{0;1,1}\, e_0 \, f_0 \, s_0 \, \sv + b_1 \, s_1 \, \sw \, \sw \quad \rangle 
\end{align*}
is a codimension three object in the full ambient space despite having four generators. However, the reducible surface with ideal $ J = \langle f_1, \, c_{1;0,0}\, e_1 \, s_1\, \sw + c_{2;0,1} \, f_0 \, s_0 \, \sv , \allowbreak \, b_{0;1,1}\, e_0 \, f_0 \, s_0 \, \sv + b_1 \, s_1 \rangle$ generated by the first, third and fourth generator of $I$ is a complete intersection, and its primary decomposition yields the associated prime components
\begin{align*}
	I , \quad \left\{
	\begin{array}{l}
		J^{(1)} = \langle f_1, s_1 , s_0 \rangle , \, J^{(2)} = \langle f_1, s_0 , \sw \rangle , \, J^{(3)} = \langle f_1 , s_1 , \sv \rangle \, , \\ [.5ex]
		J^{(4)} = \langle f_1 , \sv , \sw \rangle , \, J^{(5)} = \langle f_1 , s_1 , f_0 \rangle , \, J^{(6)} = \langle f_1 , \sw , f_0 \rangle \, ,
	\end{array}
	\right.
\end{align*}
each with multiplicity 1. All ideals except $I$ are complete intersections, hence their homology classes are obvious. In fact, because of the Stanley--Reisner ideal (\ref{eq:SR-ideal-IxA}), all the homology classes of the components $J^{(m)}$ are all 0, as at least two of their generators are not allowed to vanish simultaneously. Therefore in this specific case we have $[\gamma_I] = [\gamma_J] = F_1 \wedge ([c_{1;0,0}] + [e_1] + [s_1] + [\sw] ) \wedge ([b_1] + [s_1] + [\sw])$. From this one can also compute the class of the other surface as $[\gamma_{\bar{I}}] = ( F_1 \wedge ([b_1]+[c_{2;0,1}]) \wedge [P_T] ) - [\gamma_I]$, since $\gamma_I$ and $\gamma_{\bar{I}}$ come from the splitting of the root $F_1$.

Note that from the homology class $[\gamma_{\cal R}]$ one can compute the Cartan ($z_i$) and $U(1)$ charges ($q_i$) of the corresponding states as
\begin{align*}
	\int_{X_5} \! [\gamma_I] \wedge {\rm Ex}_i \wedge D^{({\cal B})} & = z_i \int_{\cal B}  [C_{\cal R}] \wedge D^{({\cal B})} \, ,\\
	\int_{X_5} \! [\gamma_I] \wedge \omega_i \wedge D^{({\cal B})}& = q_i \int_{\cal B} [C_{\cal R}] \wedge D^{({\cal B})} \, ,
\end{align*}
where $C_{\cal R}$ is the corresponding matter curve in the base. Applying this analysis to the above example identifies the 4-cycle $\gamma_I$ as the matter surface of a weight vector of ${\bf 3}^{\rm A}_3$ (and correspondingly $\gamma_{\bar{I}}$ as that of $\overline{\bf 3}^{\rm A}_3$).

The homology classes obtained in this way are elements of $H^{(3,3)}(X_5)$. For the study of anomalies we would like to find their possible representation as 4-cycles on the hypersurface, i.e.~as elements in $H^{(2,2)}_\text{\tiny vert} (Y_4)$. 
To this end, we make the ansatz $[\gamma] = [P_T] \wedge \sum_{i} \lambda_i \, t_i  \in H^{(3,3)}(X_5)$ for a matter surface $\gamma$, where $\{t_i\}$ are basis elements of $H^{(2,2)}_\text{\tiny vert} (Y_4)$ obtained as explained in appendix \ref{sec:calculating_flux_basis}. By equating coefficients of basis elements of  $H^{(3,3)}(X_5)$ on both sides, we can try to solve for $\lambda_i$. If there is a solution, the resulting expression $[\tilde\gamma] = \sum_i \lambda_i \, t_i$ is the desired vertical $(2,2)$-form that represents the matter surface on the fourfold.

Using the above method we have identified the homology classes of all matter surfaces except for ${\bf 1}^{(2)}$, ${\bf 1}^{(4)}$ and ${\bf 1}^{(6)}$. It was also possible to find their corresponding 4-cycle representation on the hypersurface, verifying that their classes all lie in the vertical homology. The results are listed in table \ref{tab:matter_surface_classes}. For the singlets ${\bf 1}^{(2)}$, ${\bf 1}^{(4)}$ and ${\bf 1}^{(6)}$, we have described a method in section \ref{sec:determining_singlet_classes} to determine an associated 4-cycle class $[\tilde{\gamma}]$, which gives a chiral index by $\int_{Y_4} G_4 \wedge [\tilde{\gamma}] $ leading to an anomaly free spectrum (at least for all vertical $G_4$-fluxes). We have included these classes for completeness in table \ref{tab:matter_surface_classes} as well.

\section{\texorpdfstring{\boldmath{$c_2(Y_4)$}}{c2(Y)} in the \texorpdfstring{\boldmath{${\rm I} \times {\rm A}$}}{I x A} Model}\label{app:c2}

For the analysis of the Witten anomaly \ref{sec:witten_anomaly} it is necessary to compute the second Chern class $c_2(Y_4)$ of the fourfold. By the adjunction formula
\begin{align}
	c(Y_4) = c( \{P_T \} ) = \frac{ c(X_5) }{1 + [P_T]} = \frac{1 + c_1(X_5) + c_2 (X_5) + ...}{1+ [P_T]}
\end{align}
we find for a Calabi--Yau fourfold with $[P_T] = c_1(X_5)$
\begin{align}
	c_2(Y_4) = c_2(X_5) - c_1(X_5) \, [P_T] + [P_T]^2 = c_2(X_5) \, .
\end{align}
For a fibration defined over a generic base ${\cal B}$, this can be calculated as
\begin{align}
	c(X_5) = c({\cal B}) \, \frac{\Pi_i (1 + D^{(T)}_i) }{ (1 + W_2) \, (1 + W_3)} \, ,
\end{align}
with $c({\cal B}) = c_1 ({\cal B}) + c_2({\cal B}) + ... = \overline{\cal K} + c_2({\cal B}) + ... \,$. The divisors $D^{(T)}_i$ are given by the vertices of the top defining the fibre ambient space. Because these include the exceptional divisors which sum up to the vertical divisors $W_{2,3}$ that are already accounted for in $c({\cal B})$, one has to divide out by the denominator.

Collecting all contributions of degree 2, we arrive at
\begin{align}\label{eq:second-chern-class}
\begin{split}
	c_2(Y_4) = & \, c_2(X_5)\\
	 = & \, c_2({\cal B}) + F_2 \, (2 \, E_1 + 2 \, F_1 + 3 \, F_2) - S_1 \, (2 \, E_1 + 5 \, S_1) \\
	- &\, \beta \, (E_1 - F_1 + F_2 - 2 \, S_0 + 5 \, S_1 - 3 \, U) +  \alpha \, (\beta - E_1 - 3 \, F_1 - 4 \, F_2 + 2 \, S_1 - 2 \, U) \\ 
	+&\, \overline{\cal K} \, (\alpha + \beta - E_1 + F_1 - F_2 + 2 \, S_0 + 2 \, S_1 + 3 \, U) + (\alpha - F_2 + S_0 + U) \, W_2 \\
 	+&\, ( 4 \, \alpha - E_1 + F_1 - 3 \, F_2 + 4 \, S_0 - 4 \, S_1 + 4 \, U) \, W_3 \, .
\end{split}
\end{align}

\section{All chiral models with \texorpdfstring{$|\chi_i| < 10$}{chiralities < 10}} \label{app:all_chiral models}

Here we give for completeness the other globally consistent flux configurations with $\chi({\bf 3}, {\bf 2}) = 3$ which we found with our search presented in section \ref{sec:summary_search}, with the restriction that the absolute value of the individual chiral indices of the other matter fields be smaller than 10. This translates into a corresponding number of exotic states which are vector-like with respect to the Standard Model gauge group, but chiral with respect to the extra $U(1)$ symmetry. Upon higgsing the latter, all these configurations give rise to models with three chiral generations of Standard Model matter, as discussed in more detail for the example in the main text.
Recall that our search was performed over the bases $\mathbb{P}^3$, ${\rm Bl}_1 \mathbb{P}^3$ and ${\rm Bl}_2 \mathbb{P}^3$. Only for the first two did we find any globally consistent fluxes with the property $|\chi_i| < 10$.

\subsection{\texorpdfstring{\boldmath{${\cal B} = \mathbb{P}^3$}}{B = P3}}

This base has only one independent divisor class $H$. The divisors of the non-abelian groups are therefore identified with this divisor: $W_2 = W_3 = H$.
The only fibration with a reasonable flux is for $\alpha = H, \beta = 2\,H$. The flux is for the identification $U(1)_Y = -1/2 \, U(1)_2$ and is given by
\begin{align*}
	-S_1\,E_1 - S_1\,F_1 - 3\,H\,F_1 - E_1\,F_1 - 2\,F_1^2 + H\,F_2 - 2\,F_1\,F_2 \, .
\end{align*}
The D3-tadpole of this flux is 60. The induced spectrum is:
\begin{align*}
	\begin{array}{c |c | c | c | c| c| c| c| c| c| c| c| c| c| c| c}
		{\cal R} & {\bf 2}_1 & {\bf 2}_2 & {\bf 2}_3 & {\bf 3}_1 & {\bf 3}_2 & {\bf 3}_3 & {\bf 3}_4 & {\bf 3}_5 & ({\bf 3,2}) & {\bf 1}^{(1)} & {\bf 1}^{(2)} & {\bf 1}^{(3)} & {\bf 1}^{(4)} & {\bf 1}^{(5)} & {\bf 1}^{(6)} \\ [.5ex]\hline \rule{0pt}{3ex} 
		\chi & -3 & 0 & -2 & 5 & -3 & 6 & -9 & -5 & 3 & -1 & -9 & 0 & -6 & -3 & 5
	\end{array}
\end{align*}

\subsection{\texorpdfstring{\boldmath{${\cal B} = {\rm Bl}_1 \mathbb{P}^3$}}{B = Bl1P3}}

This base has two independent divisor classes $H$ and $X$.
With the identification of the non-abelian divisors as $W_2 = X$ and $W_3 = H$, there are three fibrations with fluxes meeting the above criterion.

\subsubsection*{\boldmath{$\alpha =0$, $\beta = -2\,H$}}

\noindent For this fibration, there are two flux solutions. The first flux is for the identification $U(1)_Y = U(1)_1$ and is given by
\begin{align*}
	\frac{1}{2} (-9\,E_1\,F_2 - 3\,E_1\,H + 6\,E_1\,S_1 - 3\,F_1\,X + 12\,H\,X + 3\,S_0\,X - 3\,S_1\,X - 3\,[\sw]\,X + 6\,X^2 )
\end{align*}
with the D3-tadpole being 40.
The chiral spectrum is:
\begin{align*}
	\begin{array}{c |c | c | c | c| c| c| c| c| c| c| c| c| c| c| c}
		{\cal R} & {\bf 2}_1 & {\bf 2}_2 & {\bf 2}_3 & {\bf 3}_1 & {\bf 3}_2 & {\bf 3}_3 & {\bf 3}_4 & {\bf 3}_5 & ({\bf 3,2}) & {\bf 1}^{(1)} & {\bf 1}^{(2)} & {\bf 1}^{(3)} & {\bf 1}^{(4)} & {\bf 1}^{(5)} & {\bf 1}^{(6)} \\ [.5ex]\hline \rule{0pt}{3ex} 
		\chi & 0 & 3 & -6 & 0 & 0 & 3 & -3 & -6 & 3 & 0 & 3 & 0 & 0 & 0 & -6
	\end{array}
\end{align*}
The second flux is for $U(1)_Y = -1/2\,U(1)_2$ and is given by
\begin{align*}
	\frac{1}{2} ( -7\,E_1\,F_2 - E_1\,H + 4\,E_1\,S_1 - 3\,F_1\,X + 12\,H\,X + 3\,S_0\,X - 3\,S_1\,X - 2\,[\sw]\,X + 6\,X^2)
\end{align*}
with D3-tadpole 42. The chiral spectrum is:
\begin{align*}
	\begin{array}{c |c | c | c | c| c| c| c| c| c| c| c| c| c| c| c}
		{\cal R} & {\bf 2}_1 & {\bf 2}_2 & {\bf 2}_3 & {\bf 3}_1 & {\bf 3}_2 & {\bf 3}_3 & {\bf 3}_4 & {\bf 3}_5 & ({\bf 3,2}) & {\bf 1}^{(1)} & {\bf 1}^{(2)} & {\bf 1}^{(3)} & {\bf 1}^{(4)} & {\bf 1}^{(5)} & {\bf 1}^{(6)} \\ [.5ex]\hline \rule{0pt}{3ex} 
		\chi & 0 & 3 & -2 & -2 & 0 & 3 & -3 & -4 & 3 & -6 & -5 & 0 & -8 & 0 & -4
	\end{array}
\end{align*}

\subsubsection*{\boldmath{$\alpha= H + X$, $\beta = -H+X$}}

\noindent For this fibration, there are two flux configurations meeting our criteria. The first flux is for the identification $U(1)_Y = U(1)_1$ and is given by
\begin{align*}
	\frac{1}{2} (-9\,E_1\,F_2 - 3\,E_1\,H + 6\,E_1\,S_1 - 3\,F_1\,X + 12\,H\,X + 3\,S_0\,X - 3\,S_1\,X - 3\,[\sw]\,X + 6\,X^2 )
\end{align*}
with D3-tadpole 41.
The chiral spectrum is:
\begin{align*}
	\begin{array}{c |c | c | c | c| c| c| c| c| c| c| c| c| c| c| c}
		{\cal R} & {\bf 2}_1 & {\bf 2}_2 & {\bf 2}_3 & {\bf 3}_1 & {\bf 3}_2 & {\bf 3}_3 & {\bf 3}_4 & {\bf 3}_5 & ({\bf 3,2}) & {\bf 1}^{(1)} & {\bf 1}^{(2)} & {\bf 1}^{(3)} & {\bf 1}^{(4)} & {\bf 1}^{(5)} & {\bf 1}^{(6)} \\ [.5ex]\hline \rule{0pt}{3ex} 
		\chi & 0 & 3 & -6 & 0 & 0 & 3 & -3 & -6 & 3 & 0 & 3 & 0 & 0 & 0 & -6
	\end{array}
\end{align*}
The second flux is for $U(1)_Y = -1/2\,U(1)_2$ and is given by
\begin{align*}
	\frac{1}{2} ( -7\,E_1\,F_2 - E_1\,H + 4\,E_1\,S_1 - 3\,F_1\,X + 12\,H\,X + 3\,S_0\,X - 3\,S_1\,X - 2\,[\sw]\,X + 6\,X^2)
\end{align*}
with D3-tadpole 43. The chiral spectrum is:
\begin{align*}
	\begin{array}{c |c | c | c | c| c| c| c| c| c| c| c| c| c| c| c}
		{\cal R} & {\bf 2}_1 & {\bf 2}_2 & {\bf 2}_3 & {\bf 3}_1 & {\bf 3}_2 & {\bf 3}_3 & {\bf 3}_4 & {\bf 3}_5 & ({\bf 3,2}) & {\bf 1}^{(1)} & {\bf 1}^{(2)} & {\bf 1}^{(3)} & {\bf 1}^{(4)} & {\bf 1}^{(5)} & {\bf 1}^{(6)} \\ [.5ex]\hline \rule{0pt}{3ex} 
		\chi & 0 & 3 & -2 & -2 & 0 & 3 & -3 & -4 & 3 & -6 & -5 & 0 & -8 & 0 & -4
	\end{array}
\end{align*}

Compared to the previous fibration, the two pairs of flux configurations are formally the same, yielding the same chiral spectrum. The only quantity they differ in are the D3-tadpoles.

\subsubsection*{\boldmath{$\alpha = H+X$, $\beta = 0$}}

\noindent For this fibration, there is one flux configuration within the restrictions we set. It is determined for the identification $U(1)_Y = U(1)_1$ and given as
\begin{align*}
	\frac{1}{2} ( -3\,E_1\,F_2 + 2\,E_1\,H - E_1\,S_1 - F_1\,X + F_2\,X + S_1\,X) 
\end{align*}
with D3-tadpole 49.
The chiral spectrum is:
\begin{align*}
	\begin{array}{c |c | c | c | c| c| c| c| c| c| c| c| c| c| c| c}
		{\cal R} & {\bf 2}_1 & {\bf 2}_2 & {\bf 2}_3 & {\bf 3}_1 & {\bf 3}_2 & {\bf 3}_3 & {\bf 3}_4 & {\bf 3}_5 & ({\bf 3,2}) & {\bf 1}^{(1)} & {\bf 1}^{(2)} & {\bf 1}^{(3)} & {\bf 1}^{(4)} & {\bf 1}^{(5)} & {\bf 1}^{(6)} \\ [.5ex]\hline \rule{0pt}{3ex} 
		\chi & -2 & 1 & -2 & -1 & -1 & 1 & -2 & -3 & 3 & 1 & 2 & 0 & 0 & -1 & -3
	\end{array}
\end{align*}

\newpage

\section{D3-Tadpole on generic Base}

Here we give, for completeness, the explicit formula for the D3-brane charge $\frac{1}{2} \int_{Y_4} G_4^2$ induced by the general flux $G_4 = \sum_i z_i \, G^{z_i}_4 + G_4^{(1)}({\cal D}) + G_4^{(2)} ({\cal D}')$ as given in \eqref{eq:general_flux_basis}. On the right hand side, the threefold products of vertical divisor classes are to be understood as intersection numbers on the base $\cal B$:

\begin{align}
	&  \frac{\chi(Y_4)}{24} -n_3 = \frac{1}{2} \int_{Y_4} G_4 \wedge G_4 =  \label{eq:general_D3-tadpole} \\	
	\begin{split}
		&\frac{1}{2}\Big[( - 12\,z_2^2 - 12\,z_2\,z_3 - 24\,z_2\,z_4 - 3\,z_3^2 - 12\,z_3\,z_4 + 24\,z_4^2 + 12\,z_4\,z_5 + z_5^2)\,W_2^2\,W_3 \\
		& + (12\,z_1\,z_2 + 6\,z_1\,z_3 + 48\,z_1\,z_4 + 6\,z_1\,z_5 - 18\,z_2^2 - 24\,z_2\,z_3 - 72\,z_2\,z_4 - 3\,z_3^2 - 60\,z_3\,z_4 - 6\,z_3\,z_5 - 72\,z_4^2 \\
		& + 12\,z_4\,z_5 + 3\,z_5^2)\,W_2\,W_3^2 + (6\,z_1^2 - 6\,z_1\,z_3 + 6\,z_1\,z_5 + 6\,z_3^2 - 6\,z_3\,z_5 + 2\,z_5^2)\,W_3^3 \\
		& + ( - 36\,z_4^2 - 12\,z_4\,z_5 - z_5^2)\,W_2^2\,\overline{\cal K} + (12\,z_1\,z_2 + 6\,z_1\,z_3 - 24\,z_1\,z_4 - 6\,z_1\,z_5 + 36\,z_2^2 + 24\,z_2\,z_3 + 108\,z_2\,z_4 \\
		& + 3\,z_3^2 + 60\,z_3\,z_4 + 6\,z_3\,z_5 + 72\,z_4^2 - 24\,z_4\,z_5 - 5\,z_5^2)\,W_2\,W_3\,\overline{\cal K} \\
		& + ( - 15\,z_1^2 + 18\,z_1\,z_3 - 12\,z_1\,z_5 - 21\,z_3^2 + 12\,z_3\,z_5 - 5\,z_5^2)\,W_3^2\,\overline{\cal K} + (18\,z_4^2 + 12\,z_4\,z_5 + 2\,z_5^2)\,W_2\,\overline{\cal K}^2 \\
		& + (6\,z_1^2 + 6\,z_1\,z_3 + 6\,z_1\,z_5 + 6\,z_3^2 - 6\,z_3\,z_5 + 4\,z_5^2)\,W_3\,\overline{\cal K}^2 - z_5^2\,\overline{\cal K}^3 \\
		& + (12\,z_2\,z_3 + 6\,z_3^2 + 12\,z_3\,z_4 - 12\,z_4\,z_5 - 2\,z_5^2)\,W_2\,W_3\,\alpha \\
		& + ( - 6\,z_1\,z_3 - 6\,z_1\,z_5 + 3\,z_3^2 + 6\,z_3\,z_5 - 3\,z_5^2)\,W_3^2\,\alpha + (12\,z_4\,z_5 + 2\,z_5^2)\,W_2\,\overline{\cal K}\,\alpha \\
		& + ( - 6\,z_1\,z_3 + 6\,z_1\,z_5 - 3\,z_3^2 - 6\,z_3\,z_5 + 5\,z_5^2)\,W_3\,\overline{\cal K}\,\alpha  - 2\,z_5^2 \,\overline{\cal K}^2\,\alpha + ( - 3\,z_3^2 + z_5^2)\,W_3\,\alpha^2 \\
		&  - z_5^2\,\overline{\cal K}\,\alpha^2 + (36\,z_4^2 + 12\,z_4\,z_5 + z_5^2)\,W_2^2\,\beta + (12\,z_1\,z_2 + 6\,z_1\,z_3 + 48\,z_1\,z_4 + 6\,z_1\,z_5 - 36\,z_2\,z_4 \\
		& - 36\,z_3\,z_4 - 6\,z_3\,z_5 - 72\,z_4^2 + 24\,z_4\,z_5 + 6\,z_5^2)\,W_2\,W_3\,\beta + (3\,z_1^2 - 6\,z_1\,z_3 + 12\,z_1\,z_5 - 12\,z_3\,z_5 \\
		& + 6\,z_5^2)\,W_3^2\,\beta + ( - 24\,z_4\,z_5 - 5\,z_5^2)\,W_2\,\overline{\cal K}\,\beta + (3\,z_1^2 + 6\,z_1\,z_3 - 12\,z_1\,z_5 + 12\,z_3\,z_5 - 10\,z_5^2)\,W_3\,\overline{\cal K}\,\beta \\
		& + 4\,z_5^2\,\overline{\cal K}^2\,\beta + ( - 12\,z_4\,z_5 - 2\,z_5^2)\,W_2\,\alpha\,\beta + ( - 6\,z_1\,z_3 - 6\,z_1\,z_5 + 6\,z_3\,z_5 - 6\,z_5^2)\,W_3\,\alpha\,\beta + 5\,z_5^2 \,\overline{\cal K}\,\alpha\,\beta \\
		& + z_5^2\,\alpha^2\,\beta + ( - 18\,z_4^2 + 12\,z_4\,z_5 + 3\,z_5^2)\,W_2\,\beta^2 + ( - 3\,z_1^2 + 6\,z_1\,z_5 - 6\,z_3\,z_5 + 6\,z_5^2)\,W_3\,\beta^2 - 5\,z_5^2\,\overline{\cal K}\,\beta^2 \\
		& - 3\,z_5^2\,\alpha\,\beta^2 + 2\,z_5^2\,\beta^3 + ( - 2\,z_2 + 2\,z_3 - 8\,z_4 - 2\,z_5)\,W_2\,W_3\,{\cal D} + ( - 2\,z_1 + 4\,z_3 - 2\,z_5)\,W_3^2\,{\cal D} \\
		& + (6\,z_4 + 2\,z_5)\,W_2\,\overline{\cal K}\,{\cal D} + (4\,z_1 - 10\,z_3 + 4\,z_5)\,W_3\,\overline{\cal K}\,{\cal D}  - 2\,z_5\,\overline{\cal K}^2\,{\cal D} + ( - 2\,z_3 + 2\,z_5)\,W_3\,\alpha\,{\cal D} \\
		&  - 2\,z_5\,\overline{\cal K}\,\alpha\,{\cal D} + ( - 6\,z_4 - 2\,z_5)\,W_2\,\beta\,{\cal D} + ( - 2\,z_1 - 4\,z_5)\,W_3\,\beta\,{\cal D} + 4\,z_5\,\overline{\cal K}\,\beta\,{\cal D} + 2\,z_5\,\alpha\,\beta\,{\cal D} \\
		& - 2\,z_5\,\beta^2\,{\cal D} + 1/2\,W_2\,{\cal D}^2 + 2/3\,W_3\,{\cal D}^2 - 2\,\overline{\cal K}\,{\cal D}^2 + (4\,z_2 + 2\,z_3 + 16\,z_4 + 2\,z_5)\,W_2\,W_3\,{\cal D}' \\
		& + (4\,z_1 - 2\,z_3 + 2\,z_5)\,W_3^2\,{\cal D}' + ( - 12\,z_4 - 2\,z_5)\,W_2\,\overline{\cal K}\,{\cal D}' + ( - 8\,z_1 + 2\,z_3 - 4\,z_5)\,W_3\,\overline{\cal K}\,{\cal D}' + 2\,z_5\,\overline{\cal K}^2\,{\cal D}' \\
		& + ( - 2\,z_3 - 2\,z_5)\,W_3\,\alpha\,{\cal D}' + 2\,z_5\,\overline{\cal K}\,\alpha\,{\cal D}' + (12\,z_4 + 2\,z_5)\,W_2\,\beta\,{\cal D}' + (4\,z_1 - 6\,z_3 + 4\,z_5)\,W_3\,\beta\,{\cal D}' \\
		& - 4\,z_5\,\overline{\cal K}\,\beta\,{\cal D}'  - 2\,z_5\,\alpha\,\beta\,{\cal D}' + 2\,z_5\,\beta^2\,{\cal D}' - 2/3\,W_3\,{\cal D}\,{\cal D}' - 2\,\overline{\cal K}\,{\cal D}\,{\cal D}' + 2\,\alpha\,{\cal D}\,{\cal D}' - 2\,\beta\,{\cal D}\,{\cal D}' \\
		& + 2/3\,W_3\,{\cal D}'^2 - 4\,\overline{\cal K}\,{\cal D}'^2 + 2\,\alpha\,{\cal D}'^2 \Big].
	\end{split} \notag
\end{align}

\end{appendix}


\newpage
\bibliography{papers}  
\bibliographystyle{custom1}

\end{document}